\documentclass[sigconf]{acmart}

\newcount\draft\draft=1 % set to 0 for publication, 1 for making note with xxx
\usepackage{booktabs} % For formal tables
\usepackage{url}
\usepackage{inlinenum}
\usepackage{threeparttable}
\usepackage{multirow}
\usepackage{subfigure}
\usepackage{color}
\usepackage{xspace}
\newcommand{\doa}{\texttt{DolphinAttack}\xspace}

\begin{document}
\title{DolphinAttack: Inaudible Voice Commands} % TODO: replace with your title

\author{Guoming Zhang}
\authornote{Guoming and Chen are co-first authors.}
\affiliation{%
	\institution{Zhejiang University}
}
\email{realzgm@zju.edu.cn}

\author{Chen Yan}
\authornotemark[1]
\affiliation{%
	\institution{Zhejiang University}
}
\email{yanchen@zju.edu.cn}

\author{Xiaoyu Ji}
\authornote{Corresponding faculty authors.}
\affiliation{%
	\institution{Zhejiang University}
}
\email{xji@zju.edu.cn}

\author{Tianchen Zhang}
\affiliation{%
	\institution{Zhejiang University}
}
\email{tianchen-zhang@zju.edu.cn}

\author{Taimin Zhang} 
\affiliation{%
	\institution{Zhejiang University}
}
\email{ztm1992fly@zju.edu.cn}

\author{Wenyuan Xu}
\authornotemark[2]
\affiliation{%
	\institution{Zhejiang University}
}
\email{xuwenyuan@zju.edu.cn}

\begin{abstract}
Speech recognition (SR) systems such as Siri or Google Now have become an increasingly popular human-computer interaction method, and have turned various systems into voice controllable systems (VCS). Prior work on attacking VCS shows that the hidden voice commands that are incomprehensible to people can control the systems. Hidden voice commands, though `hidden', are nonetheless audible. In this work, we design a completely inaudible attack, \texttt{DolphinAttack}, that modulates voice commands on ultrasonic carriers (e.g., $f > 20$ kHz)  to achieve inaudibility. By leveraging the nonlinearity of the microphone circuits, the modulated low-frequency audio commands can be successfully demodulated, recovered, and more importantly interpreted by the speech recognition systems. We validate \texttt{DolphinAttack} on popular speech recognition systems, including Siri, Google Now, Samsung S Voice, Huawei HiVoice, Cortana and Alexa. By injecting a sequence of inaudible voice commands, we show a few proof-of-concept attacks, which include activating Siri to initiate a FaceTime call on iPhone, activating Google Now to switch the phone to the airplane mode, and even manipulating the navigation system in an Audi automobile. We propose hardware and software defense solutions. We validate that it is feasible to detect \texttt{DolphinAttack} by classifying the audios using supported vector machine (SVM), and suggest to re-design voice controllable systems to be resilient to inaudible voice command attacks.  

\end{abstract}

% TODO: replace this section with code generated by the tool at https://dl.acm.org/ccs.cfm
%\begin{CCSXML}
%	<ccs2012>
%	<concept>
%	<concept_id>10002978.10003001.10003003</concept_id>
%	<concept_desc>Security and privacy~Embedded systems security</concept_desc>
%	<concept_significance>500</concept_significance>
%	</concept>
%	<concept>
%	<concept_id>10002978.10003014.10003017</concept_id>
%	<concept_desc>Security and privacy~Mobile and wireless security</concept_desc>
%	<concept_significance>500</concept_significance>
%	</concept>
%	</ccs2012>
%\end{CCSXML}
%
%\ccsdesc[500]{Security and privacy~Embedded systems security}
%\ccsdesc[500]{Security and privacy~Mobile and wireless security}
% -- end of section to replace with generated code

 % TODO: replace with your keywords

\copyrightyear{2017}
\acmYear{2017}
\setcopyright{acmcopyright}
\acmConference{CCS '17}{October 30-November 3, 2017}{Dallas, TX,
	USA}\acmPrice{15.00}\acmDOI{10.1145/3133956.3134052}
\acmISBN{978-1-4503-4946-8/17/10}

\maketitle

% TODO: replace with your brilliant paper!
% \input{introduction.tex}
% \input{background.tex}
% \input{threatModel.tex}
% \input{analysis.tex}
% \input{attack.tex}
% \input{experiment.tex}
% \input{evaluation.tex}
% %\input{scenario.tex}
% \input{defense.tex}
% \input{relatedWork.tex}
% \input{conclusion.tex}

% !TEX root = CCS2017_DolphinAttack.tex
\section{Introduction}
\label{sec:introduction}

Speech recognition (SR) technologies allow machines or programs to identify spoken words and convert them into machine-readable formats. It has become an increasingly popular human-computer interaction mechanism because of its accessibility, efficiency, and recent advances in recognition accuracy. As a result, speech recognition systems have turned a wide variety of systems into voice controllable systems (VCS):   Apple Siri~\cite{Siri} and Google Now~\cite{GoogleNow} allow users to initiate phone calls by voices;  Alexa~\cite{Alexa} has enabled users to instruct an Amazon Echo to order takeouts, schedule a Uber ride, etc. As researchers devote much of their effort into improving the performance of SR systems, what is less well understood is how speech recognition and the voice controllable systems behave under intentional and sneaky attacks. % and what is the security consequence of such attacks.  

Prior work~\cite{carlini2016hidden,vaidya2015cocaine} has shown that obfuscated voice commands which are incomprehensible to human can be understood by SR systems, and thus may control the systems without being detected. Such voice commands, though `hidden', are nonetheless audible and remain conspicuous. This paper aims at examining the feasibility of the attacks that are difficult to detect, and the paper is driven by the following key questions: \textit{Can voice commands be \textbf{inaudible} to human while still being audible to devices and intelligible to speech recognition systems? Can injecting a sequence of inaudible voice commands lead to unnoticed security breaches to the voice controllable systems? } To answer these questions, we designed \textit{\doa}, an approach to inject inaudible voice commands at VCS by exploiting the ultrasound channel (i.e., $f > 20$ kHz) and the vulnerability of the underlying audio hardware. 

Inaudible voice commands may appear to be  unfeasible with the following doubts.
\begin{inlinenum}
\item \textit{How can inaudible sounds be audible to devices?} The upper bound frequency of human voices and human hearing is 20~kHz. Thus, most audio-capable devices (e.g., phones) adopt audio sampling rates lower than 44~kHz, and apply low-pass filters to eliminate signals above 20 kHz~\cite{lee2015chirp}. Previous work~\cite{vaidya2015cocaine} considers it impossible to receive voices above 20~kHz. 
\item \textit{How can inaudible sounds be intelligible to SR systems?} Even if the ultrasound is received and correctly sampled by hardware, SR systems will not recognize signals that do not match human tonal features, and therefore unable to interpret commands.
\item \textit{How can inaudible sounds cause unnoticed security breach to VCS?} The first step towards controlling VCSs is to activate them. Many VCSs (e.g., smartphones and smart home devices) implement the always-on feature that allows them to be activated by speaker-dependent wake words, i.e.,  such systems utilize voice recognition to authenticate a user. A random voice command will not pass the voice recognition.  %that are trained on a user's voice. % and can only be activated after successful voice identification, i.e., trainer's voice.
\end{inlinenum}
We solved all these problems, and we show that the \doa voice commands, though totally inaudible and therefore imperceptible to human, can be received by the audio hardware of devices, and correctly understood by speech recognition systems. We validated \doa on major speech recognition systems, including Siri, Google Now, Samsung S Voice~\cite{SVoice}, Huawei HiVoice~\cite{HiVoice}, Cortana~\cite{Cortana}, and Alexa.

Inaudible voice commands question the common design assumption that adversaries may at most try to manipulate a VCS vocally and can be detected by an alert user. Furthermore, we characterize the security consequences of such an assumption by asking the following: to what extent a sequence of  inaudible voice commands can compromise the security of VCSs. To illustrate, we show that \doa can achieve the following sneaky attacks purely by a sequence of inaudible voice commands: 

\begin{enumerate}
		%\vspace{-6pt}
\item \textit{Visiting a malicious website.} The device can open a malicious website, which can launch a drive-by-download attack or exploit a device with 0-day vulnerabilities. %, which may compromise the device and serve for future attacks. If it is a phishing website, the user may be deceived to enter important credentials.
\item \textit{Spying.} An adversary can make the victim device initiate outgoing video/phone calls, therefore getting access to the image/sound of device surroundings.
\item \textit{Injecting fake information.} An adversary may instruct the victim device to send fake text messages and emails, to publish fake online posts, to add fake events to a calendar, etc. %, which can be used against the user with social engineering.
\item \textit{Denial of service.} An adversary may inject commands to turn on the airplane mode, disconnecting all wireless communications.
\item \textit{Concealing attacks.} The screen display and voice feedback may expose the attacks. The adversary may decrease the odds by dimming the screen and lowering the volume.
%\item \textit{Transferring money.} \xxx[Yan]{Don't know yet.}
\end{enumerate}

We have tested these attacks on 16 VCS models including Apple iPhone, Google Nexus, Amazon Echo, and automobiles. Each attack is successful on at least one SR system. We believe this list is by far not comprehensive. Nevertheless, it serves as a wake-up call to reconsider what functionality and levels of human interaction shall be supported in voice controllable systems.

In summary, our contributions are listed as follows.
\begin{itemize}
\item We present \doa that can inject covert voice commands at state-of-the-art speech recognition systems by exploiting inaudible sounds and the property of audio circuits. We validate \doa on 7 popular speech recognition systems (e.g., Siri, Google Now, Alexa) across 16 common voice controllable system platforms. 
\item We show that adversaries can inject a sequence of inaudible voice commands to activate always-on system and achieve various malicious attacks. Tested attacks include launching Facetime on iPhones, playing music on an Amazon Echo and manipulating the navigation system in an Audi automobile.
%without any knowledge of the speech recognition systems can imperceptibly invoke and instruct the system, posing security and privacy threats to users.
\item We suggest both hardware-based and software-based defense strategies to alleviate the attacks, and we provide suggestions to enhance the security of voice controllable systems. 
%\item First successful attack on Siri such as call or facetime barely perceptible.
%\item We can attack n types of speech recognition system and arbitrary voice commands.
%\item Presents a new effective and convert way to communicate within different devices.
\end{itemize}

% !TEX root = CCS2017_DolphinAttack.tex
\section{Background and Threat Model}
\label{sec:background}
In this section, we introduce popular voice controllable systems, and discuss their architecture with a focus on MEMS microphones. % foundation of Dohave a brief 

\subsection{Voice Controllable System}
\begin{figure}[t]
	\includegraphics[width=0.49\textwidth]{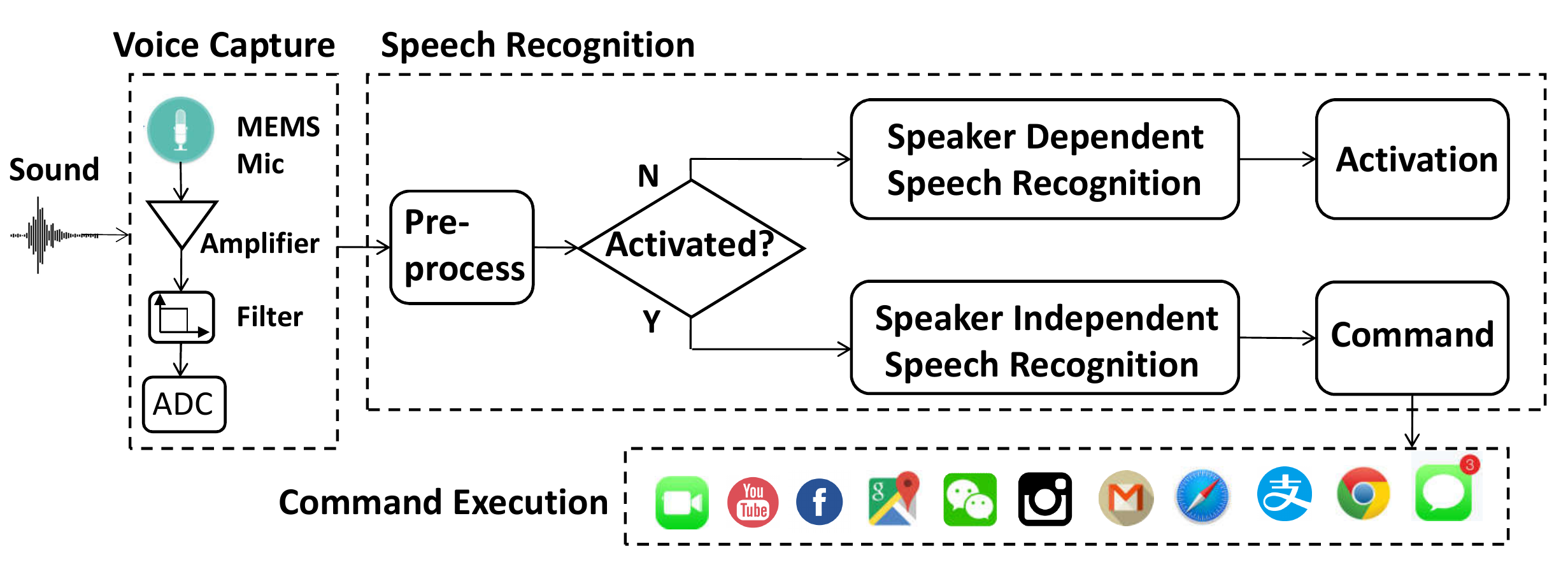}
	
%	\captionsetup{belowskip=-5pt}
	\caption{The architecture of a state-of-the-art VCS that can take voice commands as inputs and execute commands. % such as Siri, Ok, Google, e.g. 
	}
	\label{Fig:SR system}
\end{figure}

A typical voice controllable system consists of three main sub-systems: \textit{voice capture, speech recognition}, and \textit{command execution}, as illustrated in Fig.~\ref{Fig:SR system}. The voice capture subsystem records ambient voices, which are amplified, filtered, and digitized, before being passed into the speech recognition subsystem. Then, the raw captured digital signals are first pre-processed to remove frequencies that are beyond the audible sound range and to discard signal segments that contain sounds too weak to be identified. Next, the processed signals enter the speech recognition system.

Typically, a speech recognition system works in two phases: activation and recognition.  During the activation phase, the system cannot accept arbitrary voice inputs, but it waits to be activated. To activate the system, a user has to either say pre-defined wake words or press a special key.  For instance, Amazon echo takes ``Alexa'' as the activation wake word. Apple Siri can be activated by pressing and holding the home button for about one second or by ``Hey Siri'' if  the \textit{``Allow Hey Siri''} feature is enabled~\footnote{For older generation of iPhones such as iPhone 4s and iPhone 6, the ``Allow Hey Siri'' mode is only available when the device is charging.}. To recognize the wake words, the microphones continue recording ambient sounds until a voice is collected.  Then, the systems will use either speaker-dependent or speaker-independent speech recognition algorithm to recognize the voice. For instance, the Amazon Echo exploits speaker-independent algorithms and accepts  `Alexa' spoken by any one as long as the voice is clear and loud. In comparison, Apple Siri is speaker dependent. Siri requires to be trained by a user and only accepts ``Hey Siri'' from the same person. Once activated, the SR system enters the recognition phase and will typically use speaker-independent algorithms to convert voices into texts, i.e., commands in our cases. 

Note that a speaker-dependent SR is typically performed locally and a speaker-independent SR is performed via a cloud service~\cite{kasmi2015iemi}. To use the cloud service, the processed signals are sent to the servers, which will extract features (typically Mel-frequency cepstral coefficients~\cite{ittichaichareon2012speech,viikki1998cepstral,carlini2016hidden}) and recognize commands via machine learning algorithms (e.g., the Hidden Markov Models or neural networks). Finally, the commands are sent back.

Given a recognized command, the command execution system will launch the corresponding application or execute an operation. The acceptable commands and corresponding actions are system dependent and defined beforehand. Popular voice controllable systems include smartphones, wearable devices, smart home devices, and automobiles.  Smartphones allow users to perform a wide range of operation via voice commands, such as dialing a phone number, sending short messages, opening a webpage, setting the phone to the airplane mode, etc. Modern automobiles accept an elaborate set of voice commands to activate and control a few in-car features, such as GPS, the entertainment system, the environmental controls, and mobile phones.  For instance, if ``call 1234567890'' is recognized, an automobile or a smartphone may start dialing the phone number 1234567890.

Many security studies on voice controllable systems focus on attacking either the speech recognition algorithms~\cite{carlini2016hidden} or command execution environment (e.g., malware). This paper aims at the voice capturing subsystem, which will be detailed in the next subsection.

\subsection{Microphone}

\begin{figure}[t]
	\includegraphics[width=0.5\textwidth]{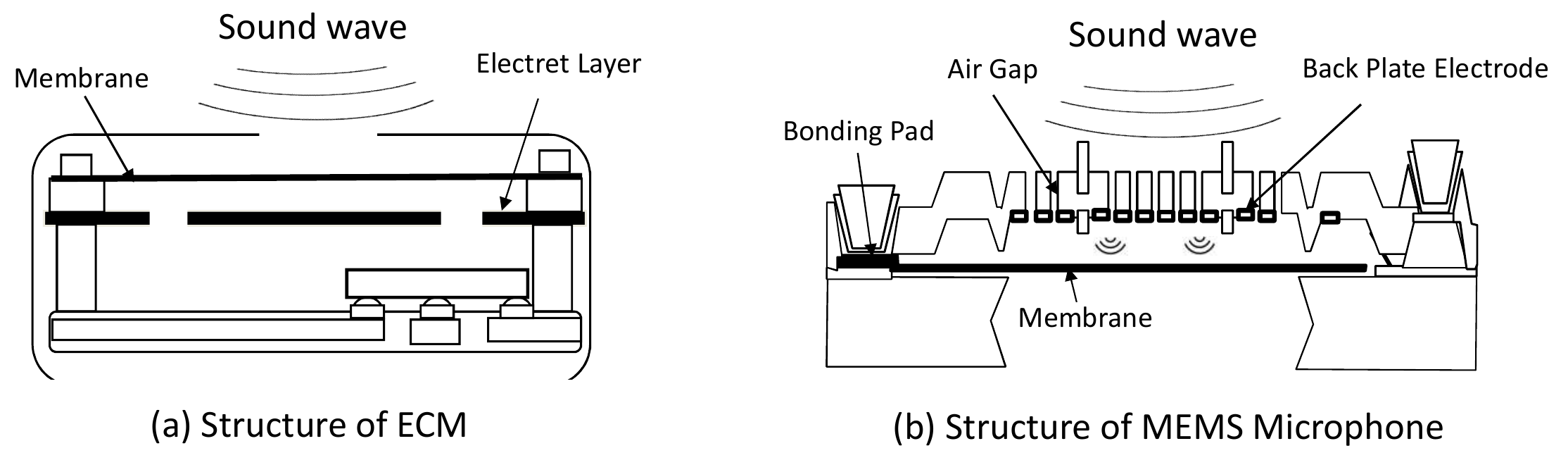}
	%	\vspace{-0.2in}
	%	\captionsetup{belowskip=-10pt}
	\caption{An illustration of the electret condenser microphone (ECM) and MEMS microphone structure.}
	%	\vspace{-4mm}
	\label{Fig:Micstructure}
\end{figure}

A voice capture subsystem records audible sounds and is mainly a microphone, which is a transducer that converts airborne acoustic waves (i.e., sounds) to electrical signals. One of the oldest and most popular types of microphones is called condenser microphones, which convert the sound waves into electrical signals via capacity changes. Both Electret Condenser Microphones (ECMs) and Micro Electro Mechanical Systems (MEMS)~\cite{Akustica1,Akustica2,STMicroelectronics1,STMicroelectronics2,Knowles1} versions are available on the market. Due to the miniature package sizes and low power consumption, MEMS microphones dominate voice controllable devices, including smartphones, wearable devices. Thus, this paper focuses mainly on MEMS microphones and will report results on ECMs briefly. Nevertheless, MEMS and ECMs work similarly. As shown in Fig.~\ref{Fig:Micstructure}(b), a MEMS microphone contains membrane (a movable plate) and a complementary perforated back-plate (a fixed plate)~\cite{MEMSmicrophone}. In the presence of a sound wave, the air pressure caused by the sound wave passes through the holes on the back-plate and reaches the membrane, which is a thin solid structure that flexes in response to the change in air pressure~\cite{wang2016windcompass}. This mechanical deformation leads to a capacitive change. Since a nearly constant charge is maintained on the capacitor, the capacitance changes will produce an AC signal. In this way, air pressure is converted into an electrical signal for further processing. Similarly, as shown in Fig.~\ref{Fig:Micstructure}(a), an ECM microphone utilizes the capacity formed by a flexible membrane and a fixed plate to record sound waves.

Designed to capture audible sounds, microphones, low-pass filters (LPFs), and ADC in the voice capture subsystem are all designed to suppress  signals out of the frequency range of audible sounds (i.e., 20 Hz to 20~kHz). According to datasheets, the sensitivity spectrum of microphones is between 20 Hz to 20 kHz, and ideally signals in any other frequency range shall be filtered.  Even if a signal higher than 20~kHz is recorded by the microphone, it is supposed to be removed by the LPF. Finally, the sampling rate of the ADC is typically 44.1~kHz, and the digitized signal's frequency is limited below 22~kHz according to the Nyquist sampling theorem.

\begin{figure}[pt]
	\includegraphics[width=0.45\textwidth]{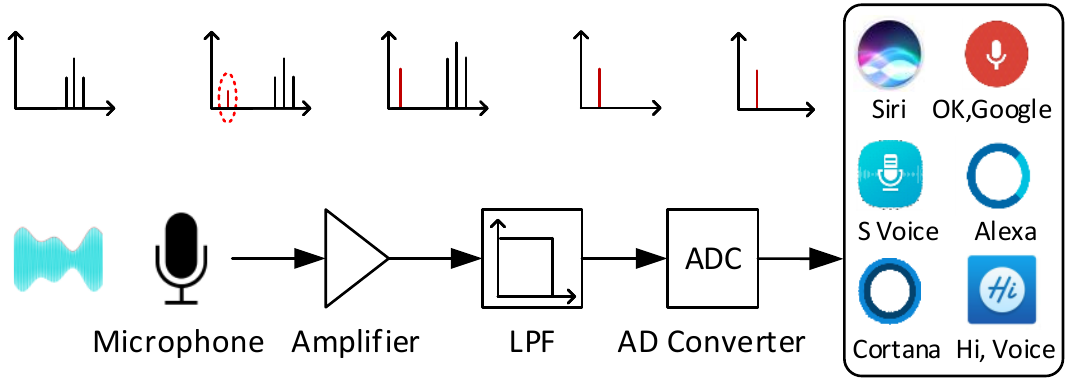}
	\caption{An illustration on the modulated tone traversing the signal pathway of a voice capture device in terms of FFT. }
	\label{fig:receiver}
\end{figure}

% !TEX root = CCS2017_DolphinAttack.tex
%\section{Threat Model}

\subsection{Threat Model}
\label{sec:threatModel}

The adversary's goal is to inject voice commands into the voice controllable systems without owners' awareness, and execute unauthenticated actions. We assume that adversaries have no direct access to the targeted device, own equipment that transmits acoustic signals, and  cannot ask the owner to perform any tasks.

\textbf{No Target Device Access.} We assume that an adversary may target at any voice controllable systems of her choices, but she has no direct access to the target devices. She cannot physically touch them, alter the device settings, or install malware. However, we assume that she is fully aware of the characteristics of the target devices. Such knowledge can be gained by first acquiring the device model and then by analyzing the device of the same model before launching attacks.

\textbf{No Owner Interaction.} We assume that the target devices may be in the owner's vicinity, but may not be in use and draw no attention (e.g., on the other side of a desk, with screen covered, or in a pocket). In addition, the device may be unattended, which can happen when the owner is temporarily away (e.g., leaving an 
Amazon Echo in a room). Alternatively, a device may be stolen, and an adversary may try every possible method to  unlock the screen. Nevertheless, the adversaries cannot ask owners to  perform any operation, such as pressing a button or unlocking the screen. %An adversary can, however, secretly record the owner as she gets close. 

\textbf{Inaudible.} Since the goal of an adversary is to inject voice commands without being detected, she will use  the sounds inaudible to human, i.e., ultrasounds ($f > 20$ kHz). Note that we did not use high-frequency sounds (18 kHz $< f <$ 20 kHz) because they are still audible to kids.

\textbf{Attacking Equipment.}  We assume that adversaries can acquire both the speakers designed for transmitting ultrasound  and commodity devices for playing audible sounds. An attacking speaker is in the vicinity of the target devices. For instance, she may secretly leave a remote controllable speaker around the victim's desk or home. Alternatively, she may be carrying a portable speaker while walking by the victim.

% !TEX root = CCS2017_DolphinAttack.tex

\section{Feasibility Analysis}
\label{sec:analysis}
The fundamental idea of \doa is (a) to modulate the low-frequency voice signal (i.e., baseband) on an ultrasonic carrier before transmitting it over the air, and (b) to demodulate the modulated voice signals with the voice capture hardware at the receiver. Since we have no control on the voice capture hardware, we have to craft the modulated signals in such a way that it can be demodulated to the baseband signal using the voice capture hardware as it is. Given that microphone modules always utilize LPF to suppress undesired high-frequency signals, the demodulation shall be accomplished prior to LPF. 

Since the signal pathway of voice capture hardware starts from a microphone, one or more amplifiers, LPF, to ADC, the potential components for demodulation are microphones and amplifiers.  We look into the principle of both to accomplish \doa.  Although electric components such as amplifiers are designed to be linear, in reality they exhibit \textbf{nonlinearity}. With this nonlinearity property, the electric component is able to create new frequencies~\cite{mixer}. Although the nonlinearity for amplifier modules is reported and utilized, it remains unknown whether a microphone, including both the ECM microphone and the MEMS one possesses such a property.

To investigate, we first theoretically model the nonlinearity for a microphone module, and then show the nonlinearity effect on real microphone modules.

\begin{figure}[pt]
	\includegraphics[width=0.485\textwidth]{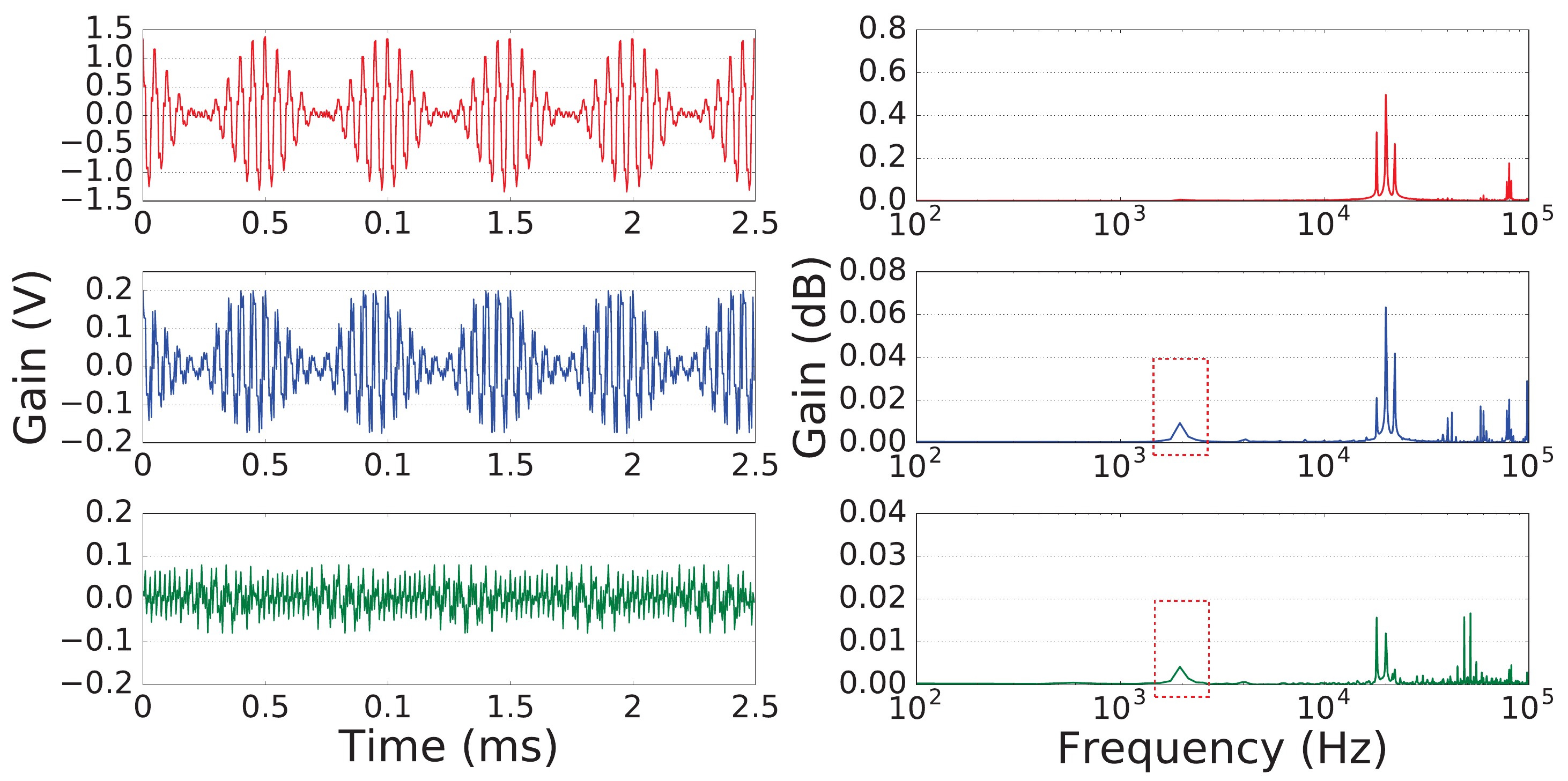}
	\caption{Evaluation of the nonlinearity effect. The time and frequency domain plots for the original signal, the output signal of the MEMS microphone, and the output signal of the ECM microphone. The presence of baseband signals at 2~kHz shows that nonlinearity can demodulate the signals.}
	\label{fig:mems}
\end{figure}

\subsection{Nonlinearity Effect Modeling}

A microphone converts mechanical sound waves into electrical signals. Essentially, a microphone can be roughly considered as a component with square-law non-linearity in the input/output signal transfer characteristics~\cite{abuelma2003analysis,chen1981comparative}. Amplifiers are known to have nonlinearity, which can produce demodulated signals in the low-frequency range~\cite{gago2007emi}.  In this paper, we study the nonlinearity of microphones and we can model it as the following.
Let the input signal be 
 $s_{in}(t)$, the output signal $s_{out}(t)$  is:
\begin{equation}
	s_{out}(t) = A s_{in}(t) + B s^2_{in}(t)
	\label{eq:amplifier}
\end{equation}
where $A$ is the gain for the input signal and $B$ is the gain for the quadratic term $s^2_{in}$. 
A linear component takes a sinusoidal input signals of frequency $f$ and outputs a sinusoidal signal with the same frequency $f$. In comparison,  
the nonlinearity of electric devices can produce harmonics and cross-products~\footnote{Harmonics are frequencies that are integer multiples of the fundamental frequency components, and cross-products are multiplicative or conjunctive combinations of harmonics and fundamental frequency components.}. Although they are typically considered undesirable distortions~\cite{kune2013ghost}, the devices with nonlinearity are able to generate new frequencies and with a crafted input signal they can downconvert the signal  as well as recover
the baseband signal.

Suppose the wanted voice control signal is $m(t)$, we choose the modulated signal on a carrier with central frequency $f_c$ to be
\begin{equation}
	s_{in}(t) = m(t)\cos(2\pi f_ct) + \cos(2\pi f_ct)
	\label{eq:input}
\end{equation}
That is, amplitude modulation is used. Without loss of generality, let $m(t)$ be a simple tone, i.e., $m(t) = \cos(2\pi f_mt)$. After applying Eq.~(\ref{eq:input}) to Eq.~(\ref{eq:amplifier}) and taking the Fourier transform, we can confirm that the output signal contains the intended frequency component $f_m$ together with the  fundamental frequency components of $s_{in}$ (i.e., $f_c - f_m$, $f_c + f_m$, and $f_c$), harmonics,  and other cross products (i.e., $f_m, 2(f_c - f_m), 2(f_c + f_m), 2f_c, 2f_c + f_m,$ and $2f_c - f_m$). After a LPF, all high-frequency components will be removed and the $f_m$ frequency component will remain, which completes the downconversion, as shown in Fig.~\ref{fig:receiver}.

\begin{figure}[pt]
	\includegraphics[width=0.48\textwidth]{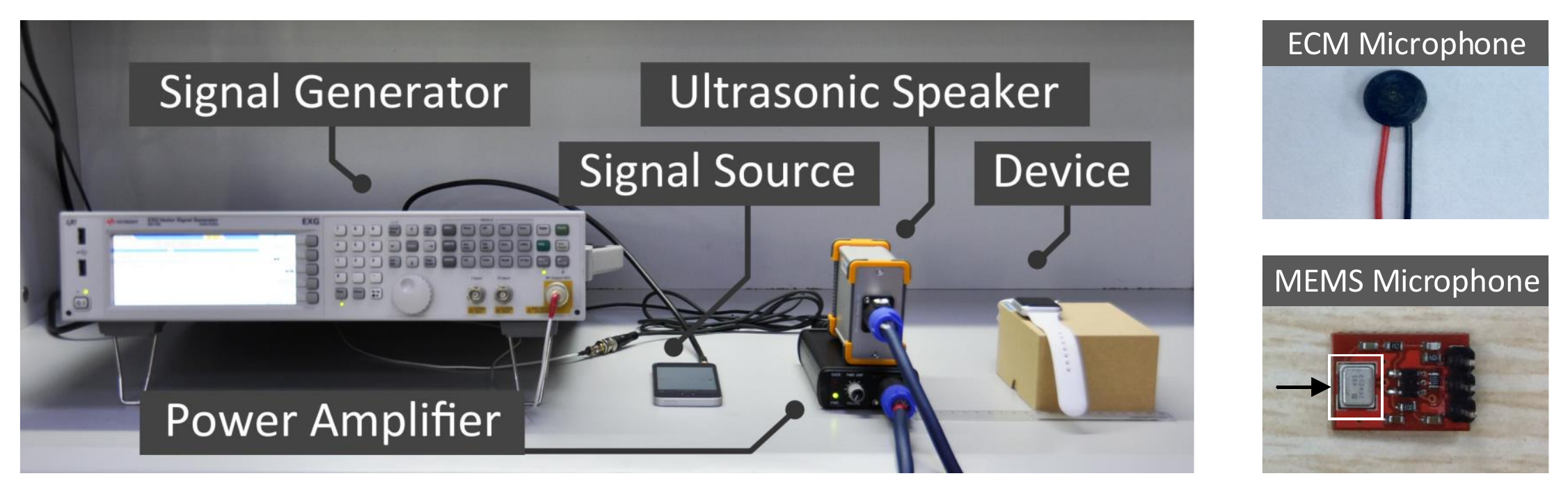}
	\caption{An illustration of the benchtop experimental setup for investigating the feasibility of receiving ultrasounds with ECM and MEMS microphones. This benchtop setup is used for validating the feasibility of attacking various VCSs as well. }
	\label{fig:setup}
\end{figure}
\subsection{Nonlinearity Effect Evaluation}
Given the theoretical calculation of the nonlinearity effect of the microphone module and its influence on the input signal after modulation, in this section, we verify the nonlinearity effect on real microphones. We test both types of microphones: ECM and MEMS microphones. 

\subsubsection{Experimental Setup}
The experimental setup is shown in Fig.~\ref{fig:setup}. We use an iPhone SE smartphone to generate a 2~kHz voice control signal, i.e. the baseband signal. The baseband signal is then inputted to a vector signal generator~\cite{keysight}, which modulates the baseband signal onto a %20~kHz 
carrier. After amplified by a power amplifier, the modulated signal is transmitted by a high-quality full-band ultrasonic speaker Vifa~\cite{vifa}.
Note that we choose the carriers ranging from 9~kHz to 20~kHz, because the signal generator cannot generate signals at the frequencies lower than 9~kHz.

On the receiver side, we test an ECM microphone that was extracted from a headphone and an ADMP401 MEMS microphone~\cite{ADMP401}. As is shown in Fig.~\ref{fig:setup}, the ADMP401 microphone module contains a preamplifier. To understand the characteristics of microphones,  we measure the signals outputted by the microphone instead of by the preamplifier.

\begin{figure}[pt]
	\includegraphics[width=0.48\textwidth]{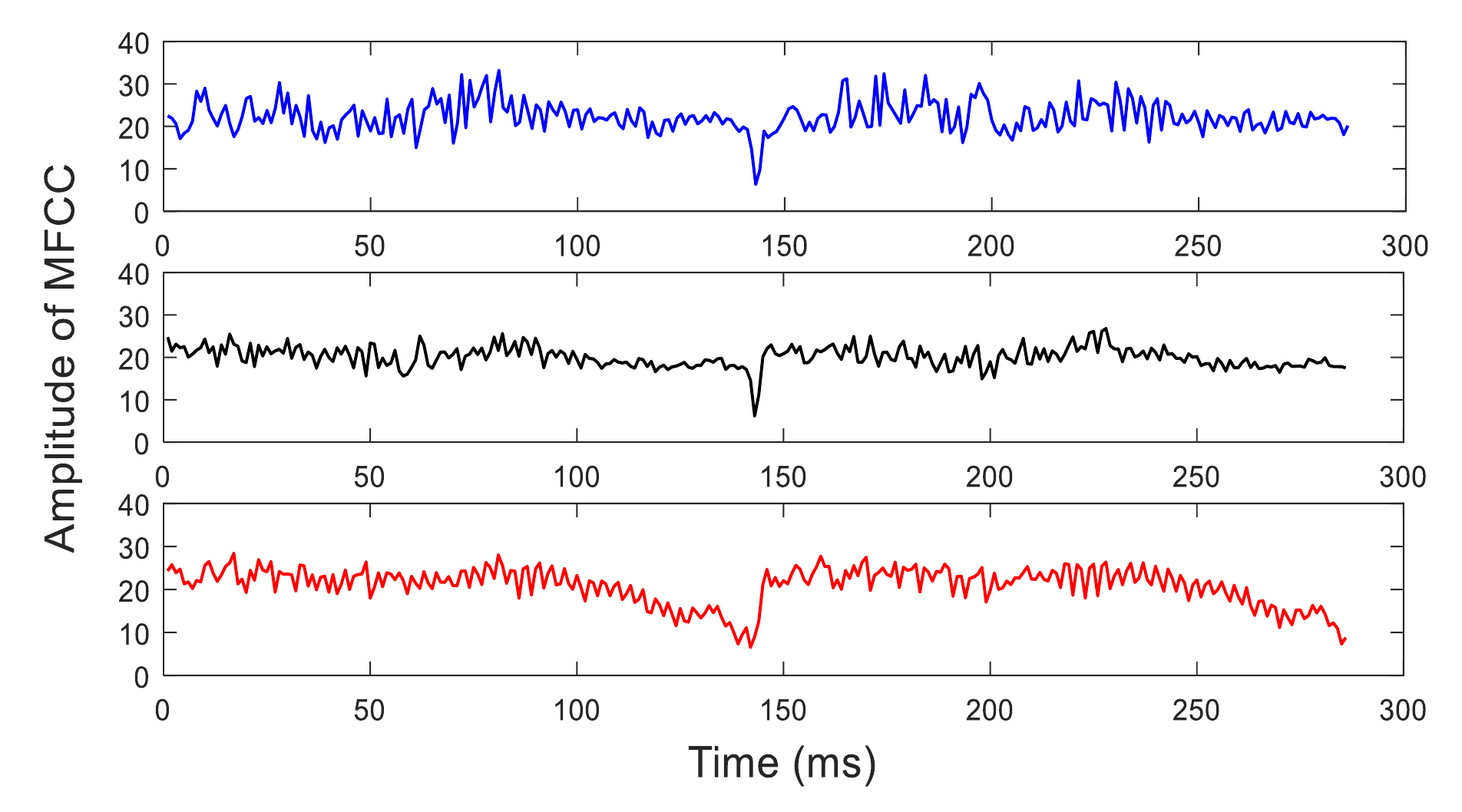}%\vspace{-3mm}
	\caption{The MFCC for three sound clips of ``Hey''. From top to bottom: the TTS generated voice, the recorded voice as the TTS voice is played in audible sounds, the recorded voice as the TTS voice is modulated to 25~kHz. }%\vspace{-3mm}
	\label{fig:mfcc}
\end{figure}

\subsubsection{Results}  We studied the nonlinearity using two types of signals: single tones and voices with multiple tones.

\textbf{Single Tone.} Fig.~\ref{fig:mems} shows the result when we use a 20~kHz carrier, which confirms that the nonlinearity of the microphone manages to demodulate the baseband signals. 
The top two figures show the original signal from the speaker in the time domain and the frequency domain, whereby the carrier frequency (20~kHz) and an upper side band as well as a lower sideband ($20\pm 2$~kHz) appear nicely.  The two figures in the second row show the output signal from the MEMS microphone and the bottom two figures depict the output signal from the ECM microphone. Even though the signals were attenuated, especially for ECM microphones, the baseband (2~kHz)  in the frequency domain for both microphones confirm the success of demodulation. Note that the frequency  domain plots include several high-frequency harmonics, which will be filtered by the LPF and shall not affect the speech recognition.

\textbf{Voices.} Even though we can demodulate a signal tone successfully, voices are a mix of numerous tones at various frequencies and it is unknown whether a demodulated voice signal remains similar to the original one. Thus, we calculated Mel-frequency cepstral coefficients (MFCC), one of the most widely used features of sounds, of three sound clips of ``Hey'': \begin{inlinenum} \item the original voice generated by a text-to-speech (TTS) engine, \item the voice recorded by a Samsung Galaxy S6 Edge as an iPhone 6 plus played the original TTS voice, and \item the voice recorded by a Samsung S6 Edge as the TTS voices are modulated and played by the full band ultrasonic speaker Vifa~\cite{vifa}. \end{inlinenum} As Fig.~\ref{fig:mfcc} shows,  the MFCC of all three cases are similar. To quantify the similarity, we calculate Mel-Cepstral Distortion (MCD) between the original one and the recorded ones, which is 3.1 for case (b) and 7.6 for case (c). MCD quantifies the distortion between two MFCCs, and the smaller the better. Typically, the two voices are considered to be acceptable to voice recognition systems if their MCD values are smaller than 8~\cite{Festvox}, and thus the result encourages us to carry out further study on \doa against voice controllable systems.

% !TEX root = CCS2017_DolphinAttack.tex
\section{Attack Design}
\label{sec:attack}

\begin{figure}[tp]
	\includegraphics[width=0.48\textwidth]{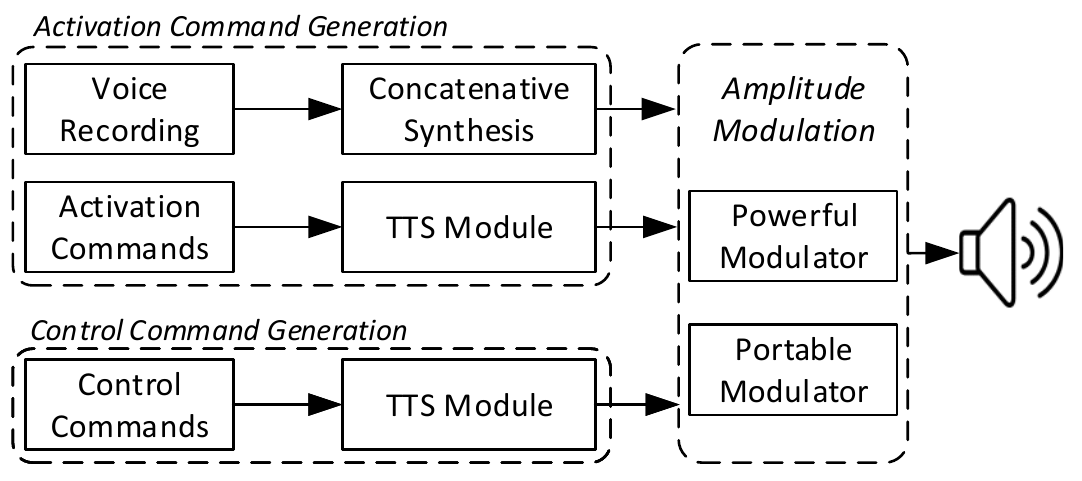}
	\caption{Architecture of the transmitter modules. The transmitter mainly includes the command generation modules and the modulation module.}
	\label{fig:AttackModule}
\end{figure}

\doa utilizes inaudible voice injection to control VCSs silently. Since attackers have little control of the VCSs, the key of a successful attack is to generate inaudible voice commands at the attacking transmitter. In particular, \doa has to generate the baseband signals of voice commands for both activation and recognition phases of the VCSs,  modulate the baseband signals such that they can be demodulated at the VCSs efficiently, and design a portable transmitter that can launch \doa anywhere. The basic building blocks of \doa are shown in Fig.~\ref{fig:AttackModule}, and we discuss these details in the following subsections. Without loss of generality, we discuss design details by using Siri as a case study, and the technology can be applied to other SR systems (e.g., Google Now, HiVoice) easily.  % that 

\subsection{Voice Command Generation}
Siri works in two phases: activation and recognition. It requires activation before accepting voice commands, and thus 
we generate two types of voice commands: activation commands and general control commands. To control a VCS, \doa has to generate activation commands before injecting general control commands.

\subsubsection{Activation Commands Generation.}

A successful activation command has to satisfy two requirements: \begin{inlinenum} \item containing the wake words ``Hey Siri'', and  \item toning to the specific voice of the user that was trained for Siri. \end{inlinenum}

Creating an activation command with both requirements is challenging, unless a user happens to speak ``Hey Siri'' when an attacker is nearby and manages to create a clear recording. In practice, an attacker can at most record arbitrary words by chances. Generating ``Hey Siri'' of the specific voice using existing speech synthesis techniques~\cite{mukhopadhyay2015all} and features extracted from the recordings is extremely different, if ever possible, because it is unclear what set of features are utilized by Siri for voice identification. As such, we design two methods to generate activation commands for two scenarios, respectively:  \begin{inlinenum} \item an attacker cannot find the owner of Siri (e.g., an attacker acquires a stolen smartphone), and \item an attacker can obtain a few recordings of the owner's voice. \end{inlinenum}

\begin{figure}[tp]
\includegraphics[width=0.48\textwidth]{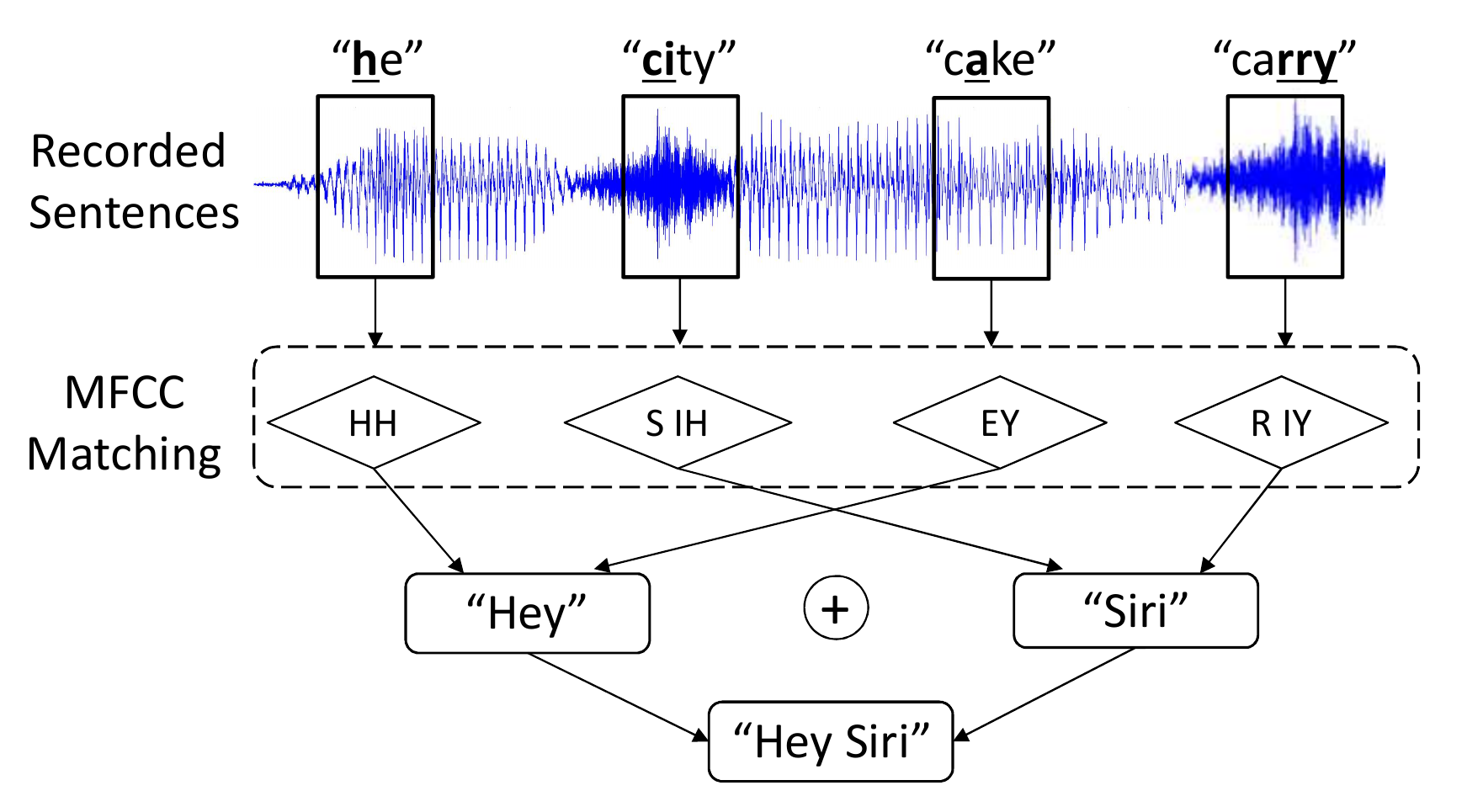}
%\vspace{-0.1in}
\caption{Concatenative synthesis of an activation command. The MFCC feature for each segment in a recorded sentence is calculated and compared with the phonemes in the activation command. After that, the matched voice segments are shuffled and concatenated in a right order.}
\label{Fig:synthesis} 	
\end{figure}

\textbf{(1) TTS-based Brute Force.} The recent advancement in TTS technique makes it easy to convert texts to voices.  Thus, even if an attacker has no chances to obtain any voice recordings from the user, she can generate a set of activation commands that contain wake words by TTS (Text to Speech) systems. This is inspired by the observation that two users with similar vocal tones can activate the other's Siri. Thus, as long as one of the activation commands in the set has a voice that is close enough to the owner, it suffices to activate Siri. In \doa, we prepare a set of activation commands with various tone and timbre with the help of existing TTS systems (summarized in Tab.~\ref{tab:TTS}), which include Selvy Speech, Baidu, Google, etc. In total, we obtain 90 types of TTS voices. We choose the Google TTS voice to train Siri and the rest for attacking. %Tab.~\ref{tab:TTS} shows the feasibility of this scheme.}

\textbf{(2) Concatenative Synthesis.} 
When an attacker can record a few words from the owner of the Siri but not necessary ``Hey Siri'', we propose to synthesize a desired voice command by searching for relevant phonemes from other words in available recordings. % For example, to obtain ``Hey, Siri'', we can separately record ``Hey'' and ``Siri'' from the two 
There are roughly 44 phonemes in English, and the wake words ``Hey Siri'' use 6 of them (i.e., HH, EY, S, IH, R, IY). %, as is shown in Tab.~\ref{tab:command}. However, 
Many words pronounce the same as ``Hey'' or ``Si'' or ``ri'', and it is possible to splice them together. For example, we can concatenate ``\underline{h}e'' and ``c\underline{a}ke'' to obtain ``Hey''. Similarly, ``Siri'' can be a combination of ``\underline{ci}ty'' and ``ca\underline{rry}''. As illustrated in Fig.~\ref{Fig:synthesis}, We
	first search for single or combined phonemes in a recorded sentence and then extracts the interested segments if a match is found. Finally, the matched phonemes are assembled.

{To evaluate the feasibility of this scheme, we conduct the following experiments. We use the Google TTS to generate ``Hey Siri'' for training the SR system,  %The experimental setup is shown in Fig.~\ref{fig:setup}.
and we generate two sets of candidate voices to synthesize ``Hey Siri'': 1. ``\underline{he}'', ``c\underline{a}ke'', ``\underline{ci}ty'', ``ca\underline{rry}''; 2. ``\underline{he} is a boy'', ``eat a c\underline{a}ke'', ``in the \underline{ci}ty'', ``\underline{re}ad after me''. After synthesizing the activation commands,  we test them on an iPhone 4S using the same experimental setup as shown in Fig.~\ref{fig:setup}. Both of the synthesized ``Hey Siri'' can activate Siri successfully.

\subsubsection{General Control Commands Generation.}
General control commands can be any commands that launch applications (e.g.,  ``call 911'', ``open www.google.com'') or configure the devices (e.g., ``turn on airplane mode''). Unlike the activation commands, an SR system does not authenticate the identities of control commands. Thus, an attacker can choose the text of any control command and utilize TTS systems to generate the command. 

\begin{table}[t]
	\centering
	\caption{The list of TTS systems used for attacking the Siri trained by the Google TTS system, and the evaluation results on activation and control commands.}
	\label{tab:TTS}
	\begin{threeparttable}
		\begin{tabular}{c|c|c|c}
			\hline
			\multirow{2}*{\textbf{TTS Systems}}& \multirow{2}*{\textbf{voice type \#}}&
			\multicolumn{2}{c}{\textbf{\# of successful types}} \\ \cline{3-4}
			&   & \textbf{Call 12..90} & \textbf{Hey Siri} \\ 
	
			\hline
			\hline
			Selvy Speech~\cite{Selvy} & 4  &  4 &  2 \\
			\hline
			Baidu~\cite{Baidu} & 1  &  1 &  0 \\
			\hline
			Sestek~\cite{Sestek} & 7  & 7 & 2  \\
			\hline
			NeoSpeech~\cite{Neospeech} & 8  & 8 & 2 \\
			\hline
			Innoetics~\cite{innoetics} & 12  & 12 & 7  \\
			\hline
			Vocalware~\cite{vocalware} & 15  & 15 & 8  \\
			\hline
			CereProc~\cite{cereproc} & 22  & 22 & 9  \\
			\hline
			Acapela~\cite{acapela} & 13  & 13 & 1 \\
			\hline
			Fromtexttospeech~\cite{fromtexttospeech} & 7  & 7 & 4 \\
			\hline								
		\end{tabular}
%		\begin{tablenotes}
%			\item[\textasteriskcentered]  We use Google TTS system to generate the training voices.
%		\end{tablenotes}
	\end{threeparttable}
	%\vspace{-5pt}
\end{table}

\subsubsection{Evaluation.}

We test both activation and control commands. Without loss of generality, we generate both activation and control commands by utilizing the TTS systems summarized in Tab.~\ref{tab:TTS}. In particular, we download two voice commands from the websites of these TTS systems: ``Hey Siri'' and ``call 1234567890''.  For activation, we use the ``Hey Siri'' from the Google TTS system to train Siri, and the rest for testing.  We play the voice commands by an iPhone 6 Plus and the benchtop devices (shown in Fig.~\ref{fig:setup}), and test on an iPhone 4S. The activation and recognition results for both commands are summarized in Tab.~\ref{tab:TTS}. The results show that the control commands from any of the TTS systems can be recognized by the SR system. 35 out of 89 types of activation commands can activate Siri, resulting in a success rate of 39\%.

\subsection{Voice Commands Modulation}

After generating the baseband signal of the voice commands,  we need to modulate them on ultrasonic carriers such that they are inaudible. To leverage the nonlinearity of microphones, \doa has to utilize amplitude modulation (AM).

\subsubsection{AM Modulation Parameters}
In AM, the amplitude of the carrier wave varies in proportion to the the baseband signal, and amplitude modulation produces a signal with its power concentrated at the carrier frequency and two adjacent sidebands, as is shown in Fig.~\ref{fig:am_modulation}. In the following, we describe how to select AM parameters in \doa.

\textbf{(1) Depth.} Modulation depth $m$  is defined as $m=M/A$ where A is the carrier amplitude, and M is the modulation amplitude, i.e., M is the peak change in the amplitude from its unmodulated value. For example, if $m=0.5$, the carrier amplitude varies by 50\% above (and below) its unmodulated level. % and 100\% if $m=1$.
 Modulation depth is directly related to the utilization of the nonlinearity effect of microphones, and our experiments show that the modulation depth is hardware dependent (detailed in Sec.~\ref{sec:experiment}). % and we elaborate the selection of this parameter with exhaustive evaluation in Sec.~\ref{sec:experiment}.

\begin{figure}[pt]
	\includegraphics[width=0.47\textwidth]{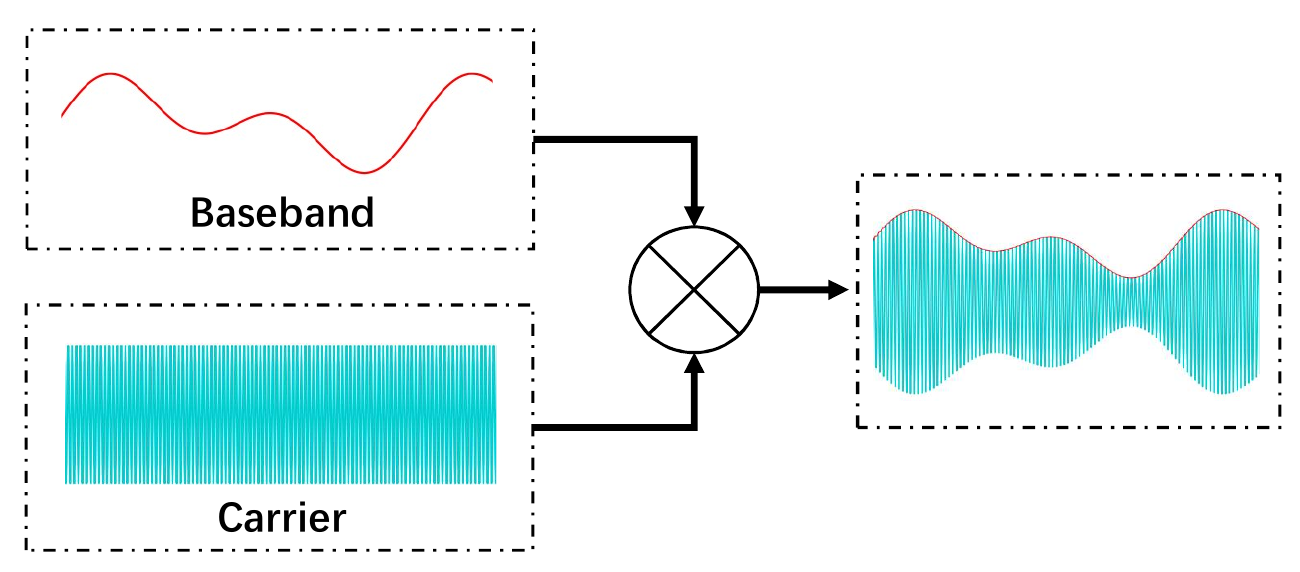}
	\caption{An illustration of modulating a voice command onto an ultrasonic carrier using AM modulation.}
	\label{fig:am_modulation}
\end{figure}

\textbf{(2) Carrier Frequency.} 

The selection of the carrier frequency depends on several factors: the frequency range of ultrasounds, the bandwidth of the baseband signal, the cut-off frequency of the low pass filter and the frequency response of the microphone on the VCS, as well as the frequency response of the attacking speaker. % and the nonlinearity effect of the microphone.
%\yc{explain why <15kHz doesn't work --- carrier is not filtered}
The lowest frequency of the modulated signal should be larger than 20 kHz to ensure inaudibility. Let the frequency range of a voice command be $w$,  the carrier frequency $f_c$ has to satisfy the condition $f_c-w > 20$ kHz. For instance, given that the bandwidth of the baseband is 6~kHz, the carrier frequency has to be larger than 26~kHz to ensure that the lowest frequency is larger than 20~kHz. 
One may consider to use the carrier that is right below 20~kHz, because these frequencies are inaudible to most people except for young kids. However, such carriers (e.g., < 20~kHz) will not be effective. This is because when the carrier frequency and lower sideband are below the cut-off frequency of the low-pass filter, they will not be filtered. Therefore, the recovered voices are different from the original signals, and the speech recognition systems will fail to recognize the commands.

Similar to many electric devices, microphones are frequency selective, e.g., the gains at various frequencies vary. For efficiency, the carrier frequency shall be the one that have the highest product of the gains at both the speaker and the VCS microphone. %An attacker should first learn the central frequency of a victim's microphone. Specifically, the frequencies with good frequency response as well as good demodulation result should be found. 
To discover the best carrier frequency, we measure the frequency response of the speaker and microphones, i.e., given the same stimulus, we measure the output magnitude at various frequencies.   Fig.~\ref{Fig:frequency response} shows the frequency response of the ADMP 401 MEMS microphone and the speaker on a Samsung Galaxy S6 Edge~\footnote{We used a professional ultrasonic microphone and speaker to assist measurement.}. The gains of the microphones and speakers do not necessarily decrease with the increase of frequencies, and thus effective carrier frequencies may not be monotonous.

\textbf{(3) Voice Selection. } Various voices map to various baseband frequency ranges. For example, a female voice typically has a wider frequency band than what a male voice has, which results in a larger probability of frequency leakage over audible frequency range, i.e., the lowest frequency of the modulated signal may be smaller than 20~kHz. Thus, if possible, a voice with a small bandwidth shall be selected to create baseband voice signals.

\begin{figure}[tp]
	\includegraphics[width=0.46\textwidth]{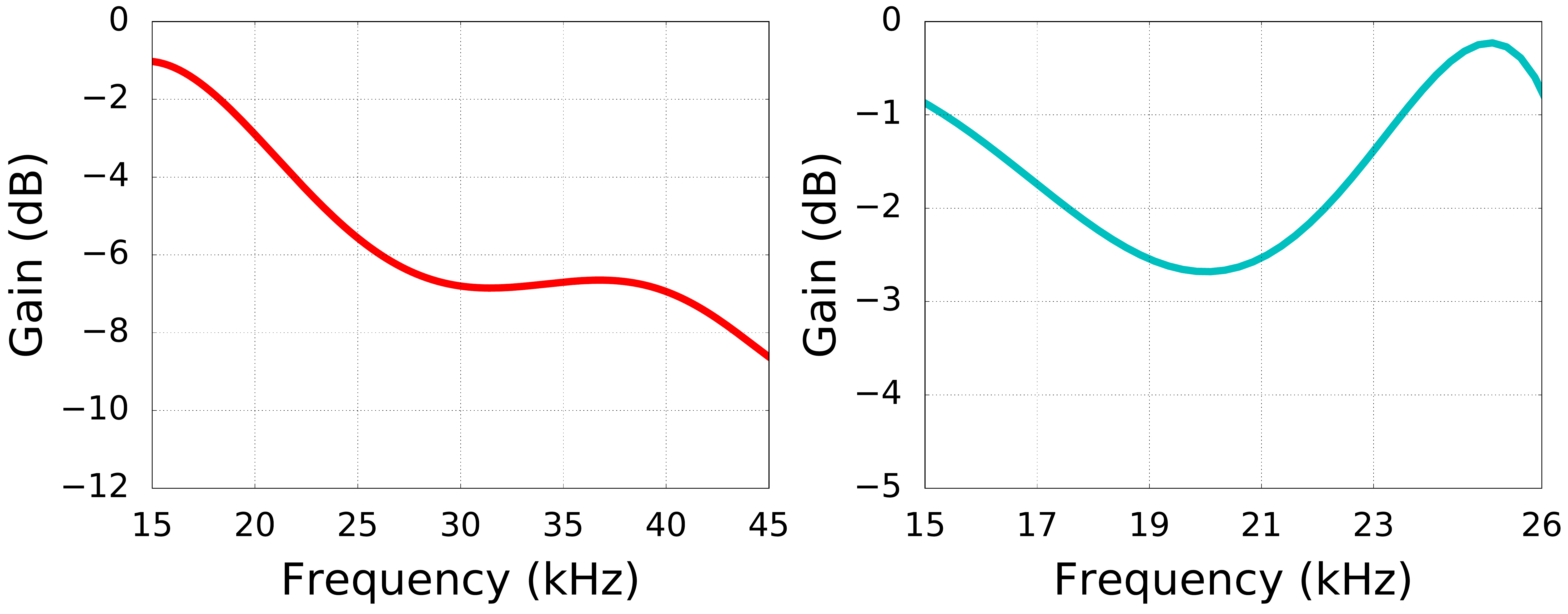}
	\caption{The frequency responses of  the ADMP401 MEMS microphone (left) and the Samsung Galaxy S6 Edge speaker (right).}
	\label{Fig:frequency response}
\end{figure}

\subsection{Voice Commands Transmitter}
We design two transmitters: \begin{inlinenum} \item a powerful transmitter that is driven by a dedicated signal generator (shown in Fig.~\ref{fig:setup}) and \item a portable transmitter that is driven by a smartphone (shown in Fig.~\ref{fig:transmitter}). \end{inlinenum} We utilize the first one to validate and quantify the extent to which \doa can accomplish various inaudible voice commands, and we use the second one to validate the feasibility of \textit{a walk-by attack}.  Both transmitters consist of three components: a signal source, a modulator, and a speaker. The signal source produces baseband signals of the original voice commands, and outputs to the modulator, which modulates the voice signal onto a carrier wave of frequency $f_c$ in forms of amplitude modulation (AM). Finally, the speaker transforms the modulated signal into acoustic waves, and note that the sampling rate of the speaker has to be larger than $2(f_c+w)$ to avoid signal aliasing.

\subsubsection{The Powerful Transmitter with A Signal Generator}
We utilize a smartphone as the signal source and the vector signal generator described in Fig.~\ref{fig:setup} as the modulator. Note that the signal generator has a sampling range of 300~MHz, much larger than ultrasonic frequencies, and can modulate signals with predefined parameters. The speaker of the powerful transmitter is a wide-band dynamic ultrasonic speaker named Vifa~\cite{vifa}.

\subsubsection{The Portable Transmitter with a Smartphone} 

The portable transmitter utilizes a smartphone to transmit the modulated signals. Since we found that the best carrier frequencies for many devices are larger than  24~kHz as is depicted in Tab.~\ref{tab:device}, a majority of smartphones cannot accomplish the task. Most smartphones support at most a 48~kHz sampling rate and can only transmit a modulated narrow-band signal with the carrier frequency of at most 24~kHz. To build a portable transmitter that works for a wide range of VCSs, we acquired a Samsung Galaxy S6 Edge, which supports a sampling rate up to 192~kHz. Unfortunately, the on-board speaker of Samsung Galaxy S6   attenuates the signal with a frequency larger than 20~kHz. To alleviate the problem, we use narrow-band ultrasonic transducers~\cite{jinci} as the speaker and add an amplifier prior to the ultrasonic transducer as shown in Fig.~\ref{fig:transmitter}. As such, the effective attack range is extended. %In this way, the modulated signal can be transmitted 
 
% !TEX root = CCS2017_DolphinAttack.tex
\section{Feasibility Experiments across VCS}
\label{sec:experiment}

\begin{figure}[tp]
	\includegraphics[width=0.45\textwidth]{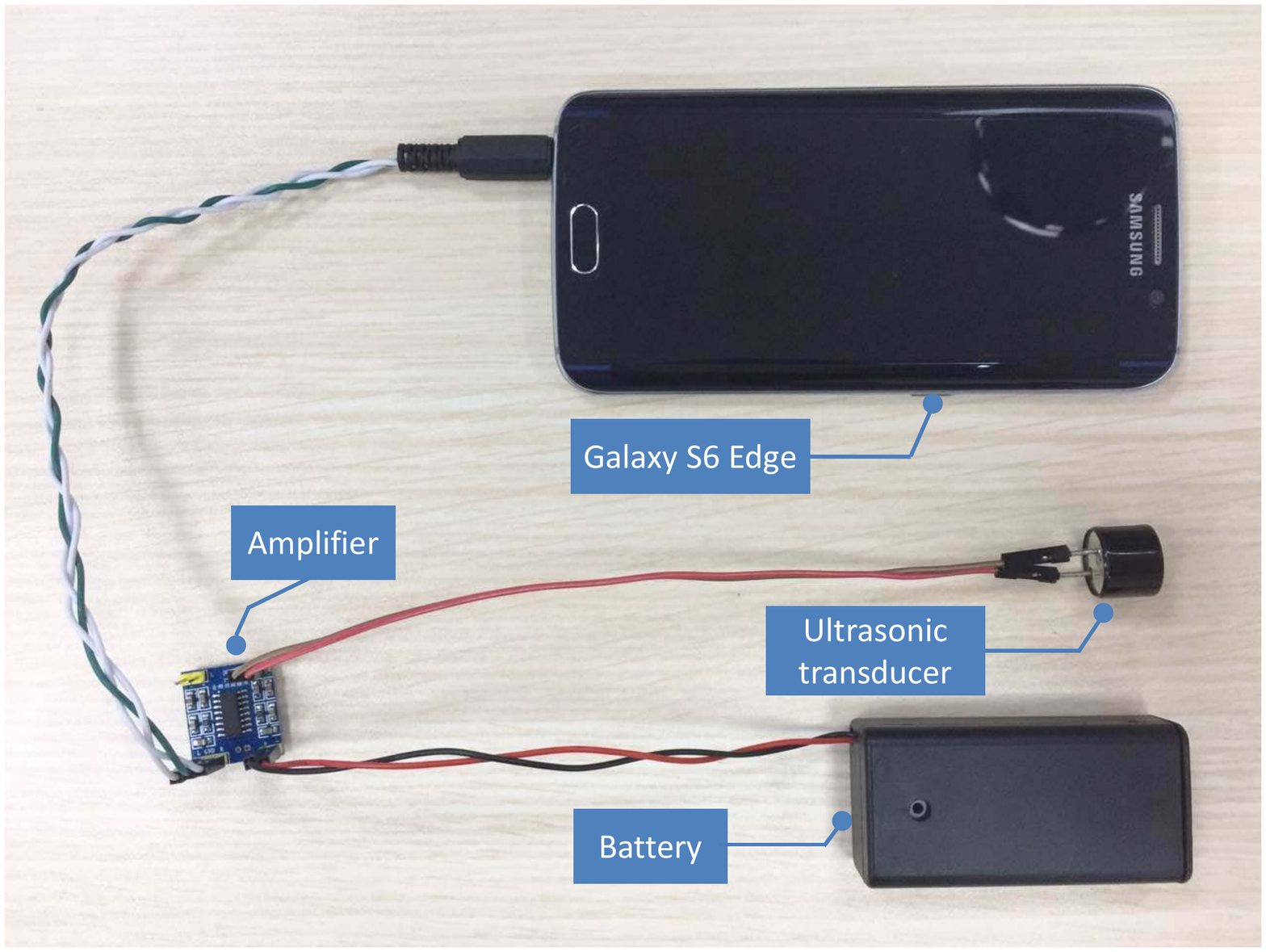}
	\caption{Portable attack implementation with a Samsung Galaxy S6 Edge smartphone, an ultrasonic transducer and a low-cost amplifier. The total price for the amplifier, the ultrasonic transducer plus the battery is less than \$3.}
	\label{fig:transmitter}
\end{figure}

We validate \doa experimentally on 16 popular voice controllable systems and 7 speech recognition systems, and seek answers to three questions: 
\begin{inlinenum}
\item Will the attacks work against different speech recognition systems on various operation systems and hardware platforms?
\item How do different software and hardware affect the performance of attacks?
\item What are the key parameters in crafting a successful attack? 
\end{inlinenum}
This section describes the experiment design, setup, and results in detail.

\subsection{System Selection}
We examine our \doa attacks on various state-of-the-art speech recognition systems and off-the-shelf VCSs, which are listed in Tab.~\ref{tab:device}.
The list does not intend to be exhaustive, but rather provides a representative set of VCSs that can be acquired for experiments with our best effort.

Our approach in selecting the target systems is twofold --- software and hardware. First of all, we select major speech recognition systems that are publicly available, e.g., Siri, Google Now, Alexa, Cortana, etc. Unlike ordinary software, SR systems (especially proprietary ones) are highly hardware and OS dependent. For example, Siri can only be found and used on Apple products; Alexa is limited to Amazon devices; Cortana runs exclusively on Windows machines. 
Nevertheless, we select and experiment on the hardware whichever the SR systems are compatible with. To explore the hardware influence on the attack performance, we examine the attacks on different hardware models running the same SR system, e.g., Siri on various generations of iPhones. 

\begin{table}[tbp]
	\centering
	\caption{The list of systems and voice commands being tested in Tab.~\ref{tab:device}.}
	%\vspace{-5pt}
	\label{tab:command}
	\begin{threeparttable}
		\begin{tabular}{l|l|l}
			\hline
			\textbf{Attack} & \textbf{Device/System} & \textbf{Command} \\
			\hline
			\hline
			Recognition & Phones \& Wearable & \textit{Call 1234567890} \\
			\hline
			Recognition & iPad & \textit{FaceTime 1234567890} \\
			\hline
			Recognition & MacBook \& Nexus 7 & \textit{Open dolphinattack.com} \\
			\hline
			Recognition & Windows PC & \textit{Turn on airplane mode} \\
			\hline
			Recognition & Amazon Echo & \textit{Open the back door} \\
			\hline
			Recognition & Vehicle (Audi Q3) & \textit{Navigation}~\tnote{\textasteriskcentered} \\
			\hline
			\hline
			%Recognition & All & \textit{How is the weather today}~\tnote{\textasteriskcentered} \\
			%\hline
			%Recognition & All & \textit{So long, and thanks for all the fish~\tnote{\textdaggerdbl}} \\
			%\hline
			Activation & Siri & \textit{Hey Siri} \\
			\hline
			Activation & Google Now & \textit{Ok Google} \\
			\hline
			Activation & Samsung S Voice & \textit{Hi Galaxy} \\
			\hline
			Activation & Huawei HiVoice & \textit{Hello Huawei}~\tnote{\textasteriskcentered} \\
			\hline
			Activation & Alexa & \textit{Alexa} \\
			\hline
		\end{tabular}
		\begin{tablenotes}
			\footnotesize
			\item[\textasteriskcentered] The command is spoken in Chinese due to the lack of English support on these devices.
			%	\item[\textdaggerdbl] From \textit{The Hitchhiker's Guide to the Galaxy} by Douglas Adams.
		\end{tablenotes}
	\end{threeparttable}
	%\vspace{-10pt}
\end{table}
%\begin{figure}[!hbt]
%\makebox[0.5\textwidth]{\framebox[5cm]{\rule{0pt}{3cm}}}
%\caption{Five by Five in Centimetres.\label{white}}
%\end{figure}

\begin{table*}
	\centering
	\caption{Experiment devices, systems, and results. \normalfont{The examined attacks include \emph{recognition} (executing control commands when the SR systems are manually activated)	 and \emph{activation} (when the SR systems are unactivated). The modulation parameters and maximum attack distances are acquired for recognition attacks in an office environment with a background noise of 55 dB SPL on average.}}
	%\vspace{-5pt}
	\label{tab:device}
	\begin{threeparttable}
		\begin{tabular}{|c|c|c|c|c|c|c|c|c|c|}
			\hline
			\multirow{2}{*}{\textbf{Manuf.}} & \multirow{2}{*}{\textbf{Model}} & \multirow{2}{*}{\textbf{OS/Ver.}} & \multirow{2}{*}{\textbf{SR System}} & \multicolumn{2}{c|}{\textbf{Attacks}} & \multicolumn{2}{c|}{\textbf{Modulation Parameters}} & \multicolumn{2}{c|}{\textbf{Max Dist. (cm)}} \\ \cline{5-10}
			& & & & \multirow{1}{0.8cm}{\textbf{Recog.}} & \multirow{1}{0.8cm}{\textbf{Activ.}} & \textbf{$f_c$ (kHz) \& [Prime $f_c$]} \textdaggerdbl & \multirow{1}{0.85cm}{\textbf{Depth}} & \multirow{1}{0.85cm}{\textbf{Recog.}} & \multirow{1}{0.85cm}{\textbf{Activ.}} \\
			\hline
			\hline
			Apple & iPhone 4s & iOS 9.3.5 & Siri & $\surd$ & $\surd$ & 20--42 [27.9] & $\geq 9\%$ & 175 & 110\\
			\hline
			Apple & iPhone 5s & iOS 10.0.2 & Siri & $\surd$ & $\surd$ & 24.1 26.2 27 29.3 [24.1] & $100\%$ & 7.5 & 10\\
			\hline
			\multirow{2}{*}{Apple} & \multirow{2}{*}{iPhone SE} & \multirow{2}{*}{iOS 10.3.1} & Siri & $\surd$ & $\surd$ & 22--28 33 [22.6] & $\geq 47\%$ & 30 & 25\\
			\cline{4-10}
			& & & Chrome & $\surd$ & N/A & 22--26 28 [22.6] & $\geq 37\%$ & 16 & N/A \\
			\hline
			Apple & iPhone SE $\dagger$ & iOS 10.3.2 & Siri & $\surd$ & $\surd$ & 21--29 31 33 [22.4] & $\geq 43\%$ & 21 & 24 \\
			\hline
			Apple & iPhone 6s \textasteriskcentered & iOS 10.2.1 & Siri & $\surd$ & $\surd$ & 26 [26] & $100\%$ & 4 & 12 \\
			\hline
			Apple & iPhone 6 Plus \textasteriskcentered & iOS 10.3.1 & Siri & $\times$ & $\surd$ & --- [24] & --- & --- & 2\\
			\hline
			%Apple & iPhone 7 & & Siri & $\surd$ & & & & &\\
			%\hline
			Apple & iPhone 7 Plus \textasteriskcentered & iOS 10.3.1 & Siri & $\surd$ & $\surd$ & 21 24-29 [25.3] & $\geq 50\%$ & 18 & 12 \\
			\hline
			Apple & watch & watchOS 3.1 & Siri & $\surd$ & $\surd$ & 20--37 [22.3] & $\geq 5\%$ & 111 & 164 \\
			\hline
			Apple & iPad mini 4 & iOS 10.2.1 & Siri & $\surd$ & $\surd$ & 22--40 [28.8] & $\geq 25\%$ & 91.6 & 50.5\\
			\hline
			Apple & MacBook & macOS Sierra & Siri & $\surd$ & N/A & 20-22 24-25 27-37 39 [22.8] & $\geq 76\%$ & 31 & N/A \\
			\hline
			LG & Nexus 5X & Android 7.1.1 & Google Now & $\surd$ & $\surd$ & 30.7 [30.7] & $100\%$ & 6 & 11 \\
			\hline
			Asus & Nexus 7 & Android 6.0.1 & Google Now & $\surd$ & $\surd$ & 24--39 [24.1] & $\geq 5\%$ & 88 & 87 \\
			\hline
			
			Samsung & Galaxy S6 edge & Android 6.0.1 & S Voice & $\surd$ & $\surd$ & 20--38 [28.4] & $\geq 17\%$ & 36.1 & 56.2 \\
			\hline
			Huawei & Honor 7 & Android 6.0 & HiVoice & $\surd$ & $\surd$ & 29--37 [29.5] & $\geq 17\%$ & 13 & 14 \\
			\hline
			%Huawei & Mate 9 & Android 7.0 & HiVoice & $\surd$ & & & & &\\
			%\hline
			%Huawei & ? & & & & & & & &\\
			%\hline
			Lenovo & ThinkPad T440p & Windows 10 & Cortana & $\surd$ & $\surd$ & 23.4--29 [23.6] & $\geq 35\%$ & 58 & 8 \\
			\hline
			%Lenovo & PC & Windows? & CMU Sphinx & $\surd$ & N/A & & & &\\
			%\hline
			% & & & iFlytek & $\surd$ & N/A & & & &\\
			%\hline
			Amazon & Echo \textasteriskcentered & 5589 & Alexa & $\surd$ & $\surd$ & 20-21 23-31 33-34 [24] & $\geq 20\%$ & 165 & 165 \\
			\hline
			Audi & Q3 & N/A & N/A & $\surd$ & N/A & 21--23 [22] & $100\%$ & 10 & N/A \\
			\hline
		\end{tabular}
		\begin{tablenotes}
			\item[\textdaggerdbl] Prime $f_c$ is the carrier wave frequency that exhibits highest baseband amplitude after demodulation. \hspace*{1.2in} --- No result
			\item[$\dagger$] Another iPhone SE with identical technical spec.
			\item[\textasteriskcentered] Experimented with the front/top microphones on devices.
		\end{tablenotes}
	\end{threeparttable}
	%\vspace{-5pt}
\end{table*}

In summary, we select VCS and SR systems that are popular on the consumer market with active users and cover various application areas and usage scenarios.  In Tab.~\ref{tab:device}, we summarize the selected VCSs for experiments, which can be classified into three categories --- personal devices (wearables, smartphones, tablets, computers), %portable personal devices (tablets, computers), 
smart home devices, and vehicles. %, which can correspond to the majority of speech recognition applications on different threat levels.

\subsection{Experiment Setup}

We test our attacks on each of the selected voice controllable system and speech recognition system with the same experiment setup and equipment, and report their behavior when injecting inaudible voice commands with three goals:
\begin{itemize}
\item Examining the feasibility of attacks.
\item Quantifying the parameters in tuning a successfully attack.
\item Measuring the attack performance.
\end{itemize}

\textbf{Equipment.} Unless specified, all experiments utilize the default experiment equipment: a powerful transmitter as shown in Fig.~\ref{fig:setup}, which consists of a smartphone as the signal source, a signal generator as the modulator, and a wide-band dynamic ultrasonic speaker named Vifa~\cite{vifa} as the speaker to play inaudible voice commands. 
Since the powerful transmitter is able to transmit signals with a wide range of carriers (from 9~kHz to 50~kHz), we use it for feasibility study. In comparison, the portable transmitter utilizes narrow-band speakers, and its transmission frequencies are limited by the available narrow-band speakers. In our case,  our portable transmitter can transmit signals at the frequencies of 23~kHz, 25~kHz, 33~kHz, 40~kHz, and 48~kHz. %, including the feasibility study of portable attacks, we use the 

\textbf{Setup.} Unless constrained by the device size, we position the selected device in front of our benchtop attack equipment at varying distances on a table, with the device microphone facing right toward the speaker. Both the device and the speaker are elevated to the same heights (i.e., 10 cm above the table) to avoid mechanical coupling.  All experiments except the one with automobiles are conducted in our laboratory with an average background noise of 55 dB SPL (sound pressure level), and we confirm that no interfering sound exists within the test frequency band (20 kHz -- 50 kHz). We play the inaudible voice commands through the powerful transmitter, and observe the results on the device screen or from device acoustic response.

Generally, multiple microphones are installed on a device to pick up voices from all directions. It is a common case that all the microphones are used in speech recognition. In our experiments, we specifically test the one that shows the best demodulation effect.

\textbf{Voice Commands.} Two categories of voice commands are prepared for two types of attacks, activation and recognition.  For those systems supporting voice activation, we try to activate them with inaudible wake word commands. To examine whether the inaudible voice commands can be correctly recognized by the speech recognition systems, we select a few English commands that are human intelligible as listed in Tab.~\ref{tab:command}. Since no commands are supported across all devices, we prepare a set of commands to cover all devices. For each command, we try two audio sources: the synthetic voices from TTS engines, and the genuine human voices spoken by the authors.

\textbf{Sound Pressure Level.} Though the sound generated for attacks are inaudible to human, we nonetheless measure the sound pressure level (SPL) in decibels using a free field measurement microphone~\cite{crysound}. The received SPL for the ultrasound is measured at 10 cm away from the Vifa~\cite{vifa} speaker and is 125 dB.

\textbf{Attacks.} In recognition attacks, the SR systems are manually activated beforehand. While in activation attacks, physical interactions with the devices are not permitted. 
% The attack experiments are yes-or-no validation, thus we use different equipments, parameters described in this paper interchangeably without further explanation~\xxx, and measure the maximum attack distance. 
The attacks are only considered successful and the distances are only recorded when the recognized texts from SR systems totally match with the attack commands.

\textbf{Modulation Parameters.} We argue that the modulation parameters may have an influence on the attack performance. We consider two factors in the amplitude modulation: the carrier wave frequency $f_c$ and the modulation depth. To quantify their influence, we place the devices 10 cm away from the wide-band ultrasonic speaker Vifa~\cite{vifa} using the Google TTS engine as the baseband audio source, and measure three values: 
\begin{inlinenum}
\item \textit{$f_c$ range} --- the range of carrier wave frequency in which recognition attacks are successful and $100\%$ accurate.
\item \textit{Prime $f_c$} --- the $f_c$ that exhibits the highest baseband~\footnote{For simplicity, the baseband signal for finding prime $f_c$ is a 400 Hz single tone which resides in human voice frequency.} amplitude after demodulation.
\item \textit{AM depth} --- the modulation depth at the prime $f_c$ when recognition attacks are successfully and $100\%$ accurate. 
\end{inlinenum}

\subsection{Feasibility Results}
Tab.~\ref{tab:device} summarizes the experiment results. From Tab.~\ref{tab:device}, we can conclude that \doa works with nearly all of the examined SR systems and devices. In particular, the inaudible voice commands can be correctly interpreted by the SR systems on all the tested hardware, and the activation is successful on all VCSs that require activation.  The results, however, do show that devices and systems require various parameters to accomplish the same attack effect. We discuss our findings as follows.

\begin{figure}[tp]
	\centering
	\includegraphics[width=0.48\textwidth]{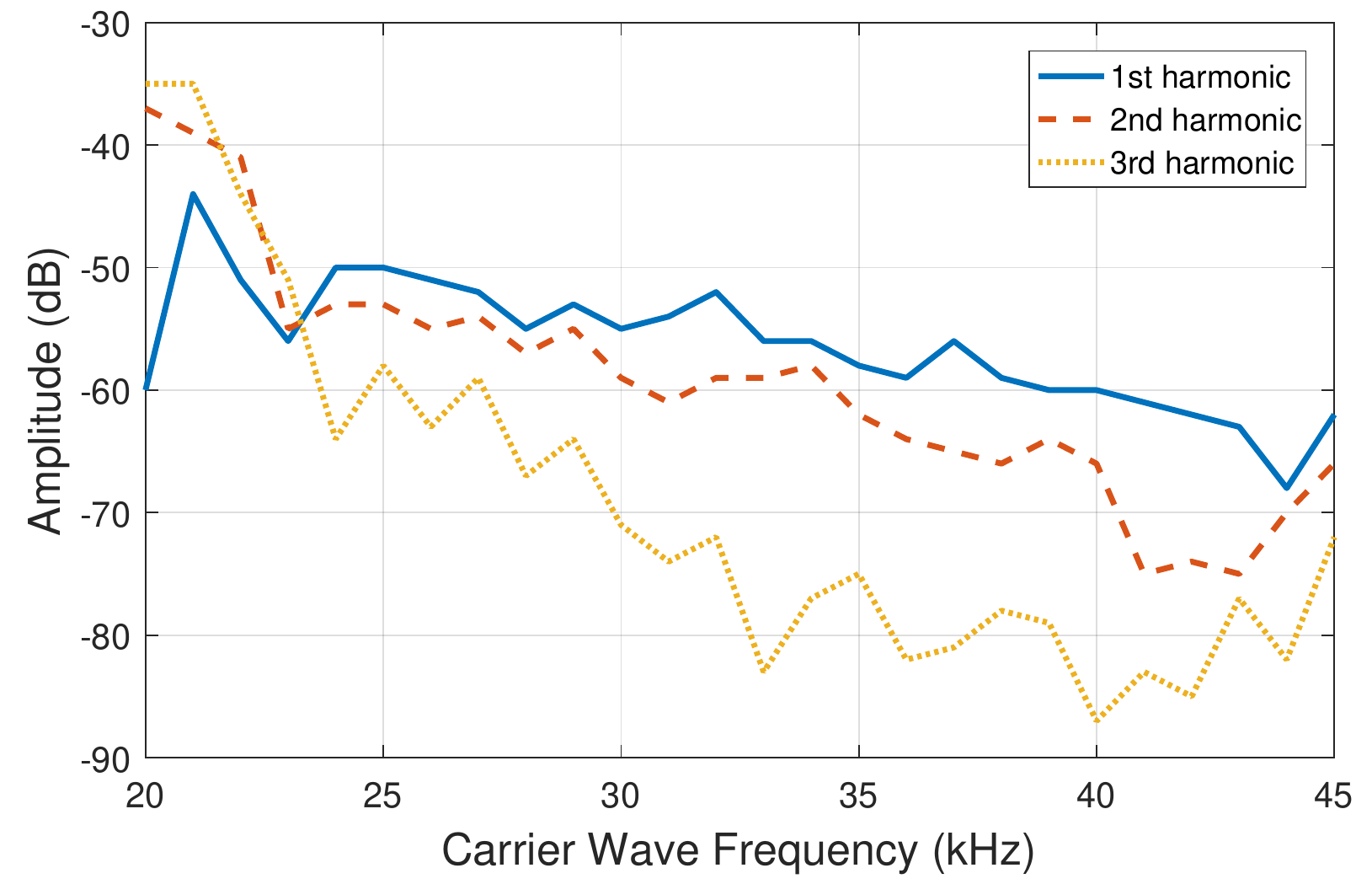}
	%	\captionsetup{belowskip=-10pt}
	\caption{Amplitude of the demodulated 400 Hz baseband signal (1st harmonic) and its higher order harmonics on Nexus 7, with varying carrier wave frequency $f_c$.}
	\label{fig:fc}
\end{figure}
\textbf{Hardware Dependence.} The basic principle of \doa is to inject inaudible voice commands before digitization components. Therefore, the feasibility of \doa depends heavily on the audio hardware rather than the speech recognition systems. For example, various devices from the same manufacturer running Siri show the great variance in the attack success rate, the maximum attack distance, and modulation parameters. This is because various models adopt different hardware (e.g., microphones, amplifiers, filters), which lead to variation in the digitized audios that are input to the same SR system. Our experiment on two identical devices (iPhone SE) exhibits similar attack parameters and results. Thus, it is feasible for an adversary to study the hardware beforehand to achieve satisfying attack results. %, and our experiments can serve as a reference.

\textbf{SR System Dependence.} We find that various SR systems may handle the same audios differently. We tested the voice search in Google Chrome running on an iPhone SE. The results in Table~\ref{tab:device} show that the $f_c$ range of Google Chrome  overlaps with the $f_c$ range in Siri experiment, which suggests that our attacks are hardware dependent. However, the differences in $f_c$, AM depth, and recognition distances are resulted from the SR systems.

\textbf{Recognition versus Activation.} Various devices and SR systems can react differently to recognition and activation attacks in terms of the attack distance. 
For some devices (8 devices), activation attacks can be achieved at a larger distance than recognition attacks, while for other devices (6 devices), the effective range of successful activation attacks is smaller than the recognition attacks. In addition, we observe that for many of the devices, appending the activation commands (e.g., ``Hey Siri'') before the control commands can increase the probability for correct recognition, possibly because the activation commands are trained specially by the SR systems to be recognized in the always-on mode.

\textbf{Commands Matter.} The length and content of a voice command can influence the success rate and the maximum distance of attacks. We are rigorous in the experiments by demanding every single word within a command to be correctly recognized, though this may be unnecessary for some commands. For instance, ``Call/FaceTime 1234567890'' and ``Open dolphinattack.com'' is harder to be recognized than ``Turn on airplane mode'' or ``How's the weather today?''. In the former scenarios, both the execution words ``call'', ``open'' and the content (number, url) have to be correctly recognized. However, for the latter scenarios, only recognizing key words such as ``airplane'' and ``weather'' shall be enough for executing the original commands. The attack performance can be improved if the attack command is short and common to SR systems.

\begin{figure}[tb]
	\centering
	\includegraphics[width=0.48\textwidth]{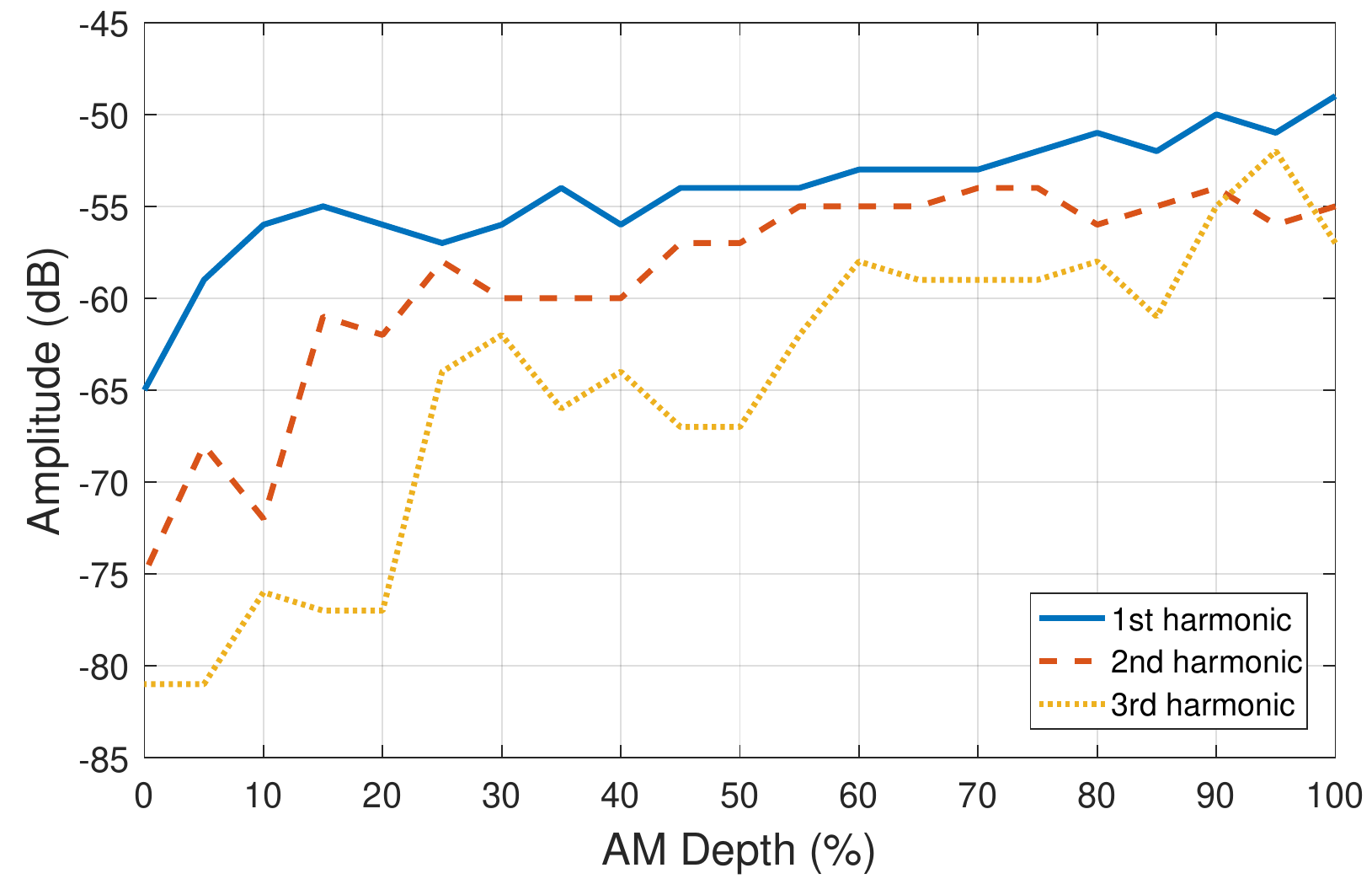}
	%	\captionsetup{belowskip=-10pt}
	\caption{Amplitude of the demodulated 400 Hz baseband signal (1st harmonic) and its higher order harmonics on Nexus 7, with varying modulation depth.}
	\label{fig:amDepth}
\end{figure}

\textbf{Carrier Wave Frequency.} $f_c$ is the dominant factor that affects the attack success rate, and it also shows great variation across devices. For some devices, the $f_c$ range within which recognition attacks are successful can be as wide as 20--42 kHz (e.g., iPhone 4s), or as narrow as a few single frequency points (e.g., iPhone 5s). We attribute this diversity to the difference of frequency response and frequency selectivity for these microphones as well as the nonlinearity of audio processing circuits. %As the experiments go, we realize that there are more reasons that influence the attacks besides microphone's frequency response. 

For instance, the $f_c$ range of Nexus 7 is from 24 to 39~kHz, which can be explained from two aspects. The $f_c$ is no higher than 39~kHz because the frequency response of the Vifa speaker over 39~kHz is low and the one of Nexus 7 microphone is low as well. Thus, in combination, a carrier higher than 39~kHz is no longer efficient enough to inject inaudible voice commands. The $f_c$ cannot be smaller than 24~kHz because of the nonlinearity of the microphone frequency response.  We observe that the inaudible voice commands become unacceptable to SR systems when the amplitude of the harmonics of the baseband are larger than the one of baseband. For instance, given the baseband of a 400~Hz tone, we measure the demodulated signal (i.e., the 400 Hz baseband) on a Nexus 7, and observe harmonics at 800 Hz (2nd harmonic), 1200 Hz (3rd harmonic) and even higher, which are possibly caused by the nonlinearity of audio processing circuits.  As shown in Fig.~\ref{fig:fc}, when the $f_c$ is less than 23~kHz,  %all harmonics weaken, which corresponds to the drop of microphone's frequency response, and explains why the attack fails when $f_c > 39$ kHz in Tab.~\ref{tab:device}. However, from our experimental results, $f_c < 24$ kHz also fails even though the demodulated signals are strong in Fig.~\ref{fig:fc}. Notice that 
the 2nd and 3rd harmonics are stronger than the 1st harmonic, which will distort the baseband signal and make it hard for SR systems to recognize. The \textit{Prime $f_c$} that leads to the best attack performance, is the frequency that exhibits both a high baseband signal and low harmonics. On Nexus 7, the \textit{Prime $f_c$} is 24.1 kHz.

\textbf{Modulation Depth.} Modulation depth affects the amplitude of demodulated baseband signal and its harmonics, as shown in Fig.~\ref{fig:amDepth}. As the modulation depth gradually increases from 0 to 100\%, the demodulated signals become stronger, which in turn increase the SNR and the attack success rate, with a few exceptions (e.g., when the harmonics distort the baseband signal more than the cases of a lower AM depth). We report the minimum depth for successful recognition attacks on each device in Tab.~\ref{tab:device}.

\textbf{Attack Distance.} The attack distances vary from 2 cm to a maximum value of 175 cm and show a great variation across devices. Notably, the maximum distance that we can achieve for both attacks is 165 cm on Amazon Echo. We argue that the distance can be increased with the equipment that can generate a sound with higher pressure levels and exhibit better acoustic directionality, or by using shorter and more recognizable commands.

\textbf{Efforts and Challenges.} We faced challenges in conducting the above experiments. Apart from acquiring the devices, measuring each parameter is time-consuming and labor-intensive due to the lack of audio measurement feedback interface. For example, to measure the \textit{Prime $f_c$}, we analyze the demodulation results on various devices using audio spectrum analyzing software on different platforms: iOS~\cite{ultrasonicAnalyzer}, macOS~\cite{ispectrum}, Android~\cite{spectroid}, and Windows~\cite{arta}. For devices not supporting installing spectrum software such as Apple watch and Amazon Echo, we utilize the calling and command log playback function, and measure the audio on another relaying device.

\subsection{Summary}
We summarize our experiments as follows.
\begin{enumerate}
	\item We validated recognition and activation attacks across 16 various devices and 7 speech recognition systems, and succeeded on nearly all of them.
	\item We measured the attack performance on all devices, and some of them suffice for real attacks in daily scenarios. For instance, we can launch \doa from almost 2 meters away against an iPhone 4s and Amazon Echo.
	\item We measured, examined, and discussed the parameters involved in the attack performance, including SR systems, device hardware, voice commands, $f_c$, AM depth, etc.
\end{enumerate}

% !TEX root = CCS2017_DolphinAttack.tex
\section{Impact Quantification}
\label{sec:evaluation}

\begin{figure}[tp]
	\includegraphics[width=0.48\textwidth]{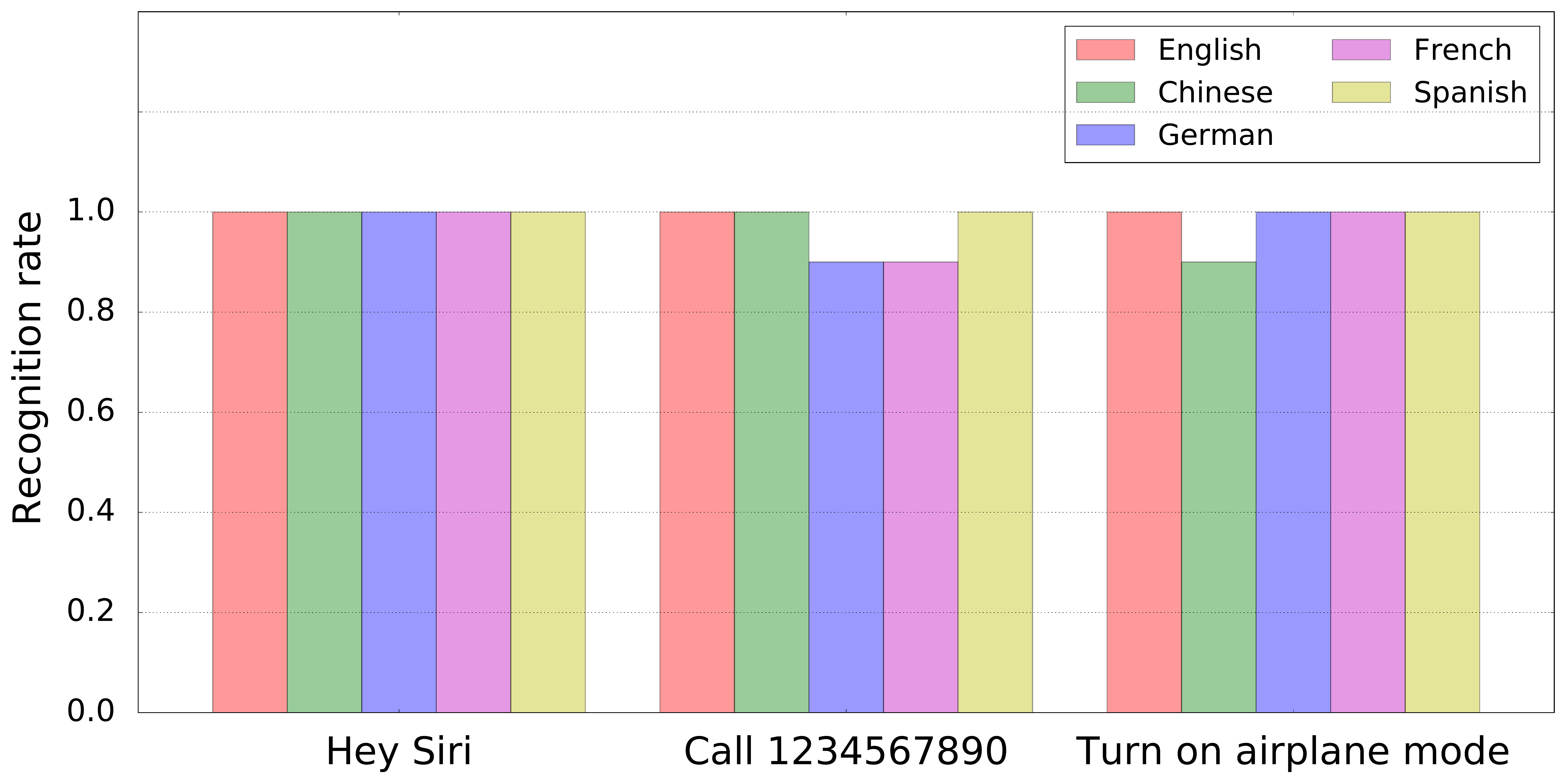}
	\caption{The recognition rates of voice commands in five languages.} 
	\label{Fig: language}	
\end{figure}

In this section, we evaluate the performance of \doa in terms of languages, background noises, sound pressure levels, and attack distances using the powerful transmitter (i.e., the benchtop setup shown in Fig.~\ref{fig:setup}). In addition, we evaluate the effectiveness of walk-by attacks using the portable devices.

\subsection{Influence of Languages}
To examine the effectiveness of \doa with regard to languages, we select three voice commands in five languages. The voice commands include an activation command (``Hey Siri'') and two control commands (``Call 1234567890'' and ``Turn on airplane mode''), which represent three attacks against SR systems: activating SR systems, initiating to spy on the user, and denial of service attacks. Each voice command is tested in English, Chinese, German, French, and Spanish, respectively.

We launch \doa against an Apple watch that is paired with an iPhone 6 Plus running iOS 10.3.1. For each voice command in each language, we repeat it for 10 times and calculate the average success rate. The distance is set to 20 cm, the measured background noise is 55 dB. We exploit a 25 kHz carrier frequency and 100\% AM depth.

Fig.~\ref{Fig: language} shows the recognition results of the three voice commands in the given languages. As we can see that the recognition rate of various languages and voice commands are almost the same. In particular, the recognition rate of all the voice commands in English and Spanish is 100\%, and the average recognition rate of the three voice commands across all languages are 100\%, 96\%, 98\%, respectively. Moreover, the recognition rate for activation (i.e., ``Hey Siri'') is higher than the one of control commands ``Call 1234567890'' and ``Turn on airplane mode''). This is because the length of the activation command is shorter than the control commands. In any case, the results show that our approach is effective for various languages and voice commands.

\begin{table}[tp]
	\centering
	\caption{The impact of background noises for sentence recognition evaluated with an Apple watch.}
	\label{tab:scene}
	\begin{tabular}{|c|c|c|c|}
		\hline
		\multirow{2}*{\textbf{Scene}}& \multirow{2}*{\textbf{Noises (dB)}}&
		\multicolumn{2}{c|}{\textbf{Recognition rates}} \\ \cline{3-4}
		&   & \textbf{Hey Siri} & \textbf{Turn on airplane mode} \\ 
		\hline \hline
		Office  & 55--65   & 100\% & 100\%    \\ \hline
		Cafe & 65--75  & 100\% &   80\% \\ \hline
		Street & 75--85  & 90\%  &  30\%  \\ \hline
	\end{tabular}
\end{table}

\begin{figure*}[tb]
	\centering
	%\vspace{-3pt}
	\subfigure[The recognition rates of the Galaxy S6 Edge]{
		\includegraphics[width=0.46\textwidth]{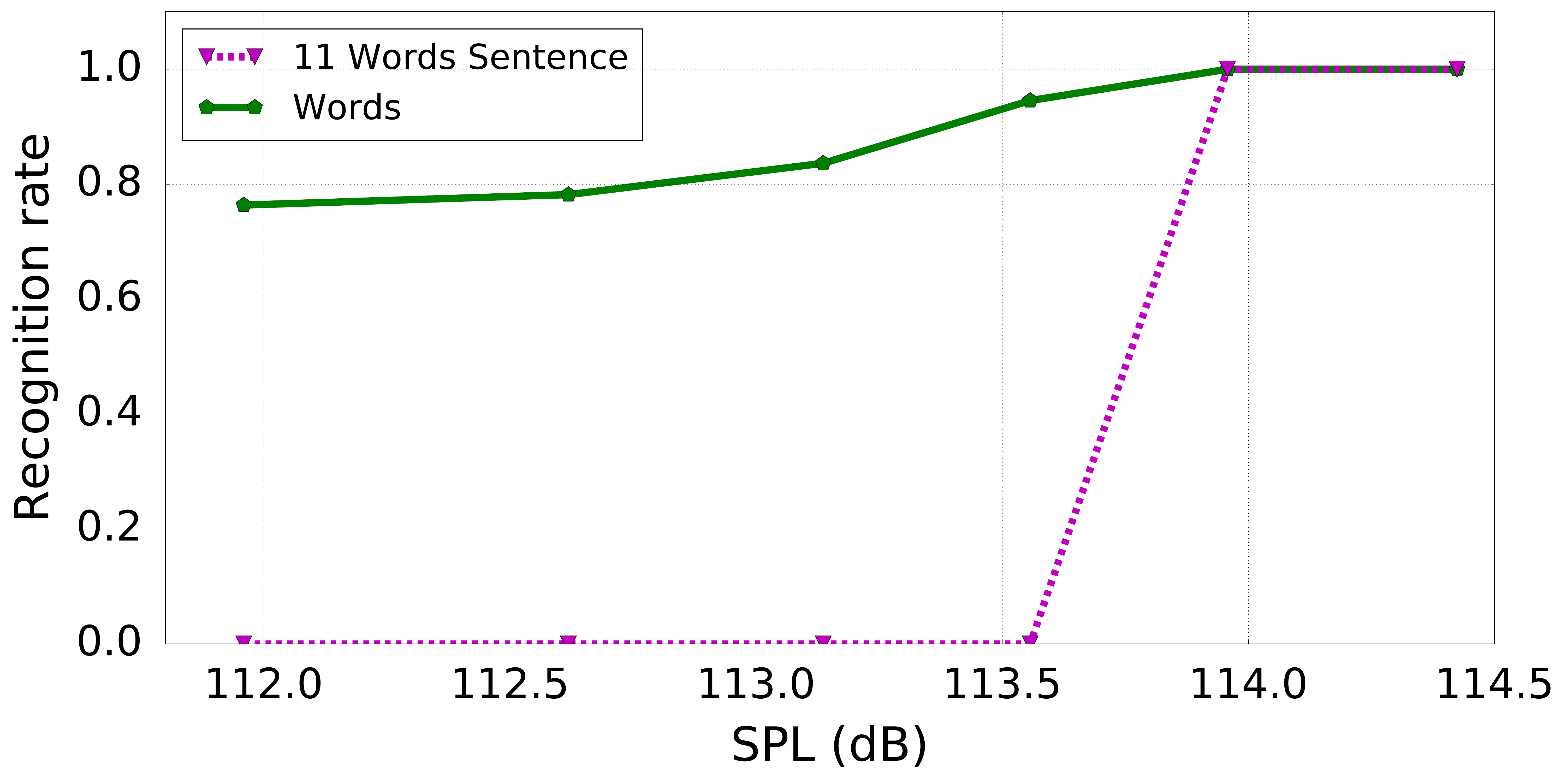}}	
	\subfigure[The recognition rates of the Apple watch]{
		\includegraphics[width=0.46\textwidth]{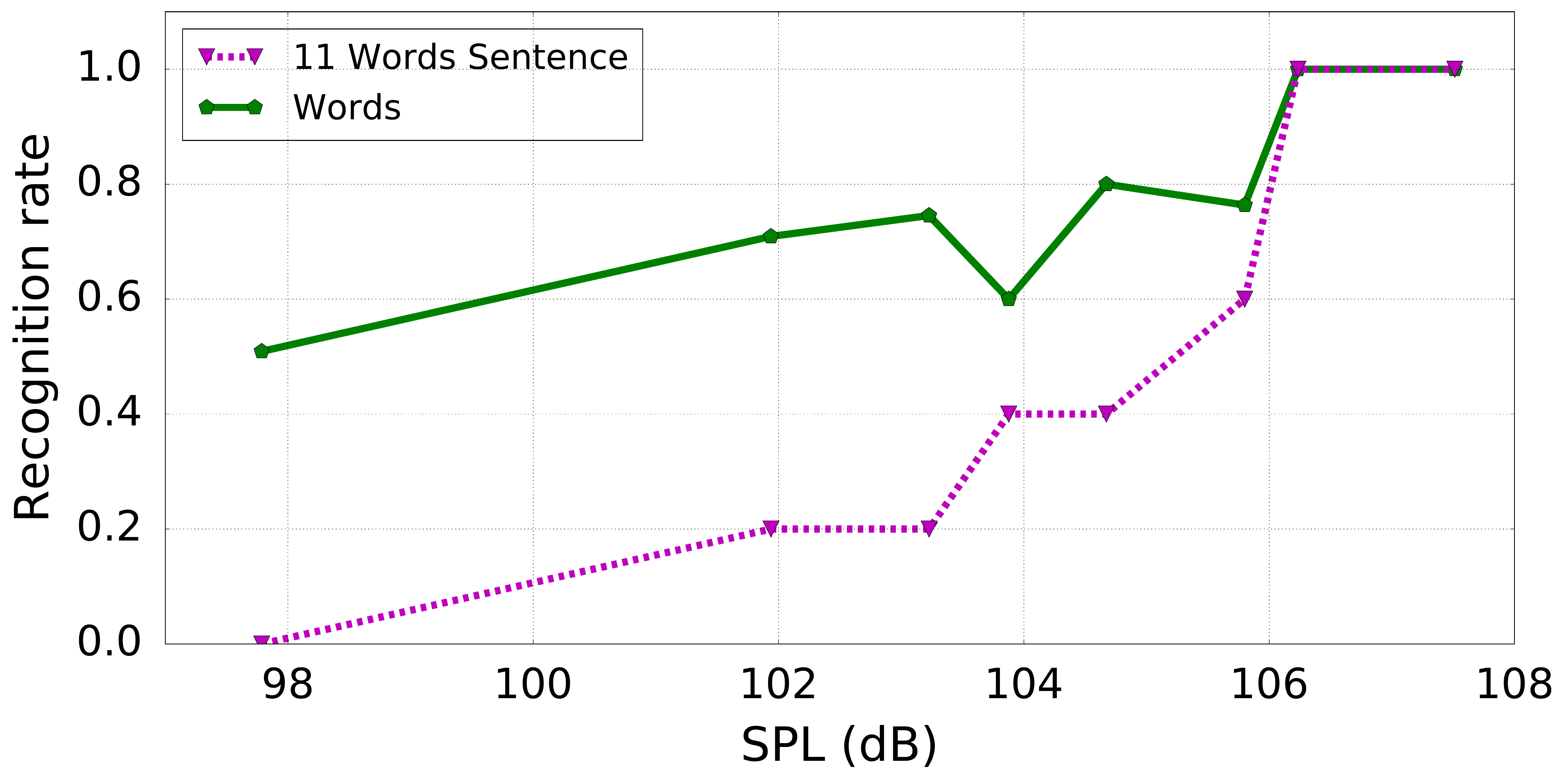}}	
	%\vspace{-0.1in}
	\caption{The impact of sound pressure levels on the recognition rates for two portable devices.}
	\label{Fig:SPL}
\end{figure*}
\begin{figure*}[tb]
	\centering
	%\vspace{-3pt}
	\subfigure[The recognition rates of the Galaxy S6 Edge]{
		\includegraphics[width=0.46\textwidth]{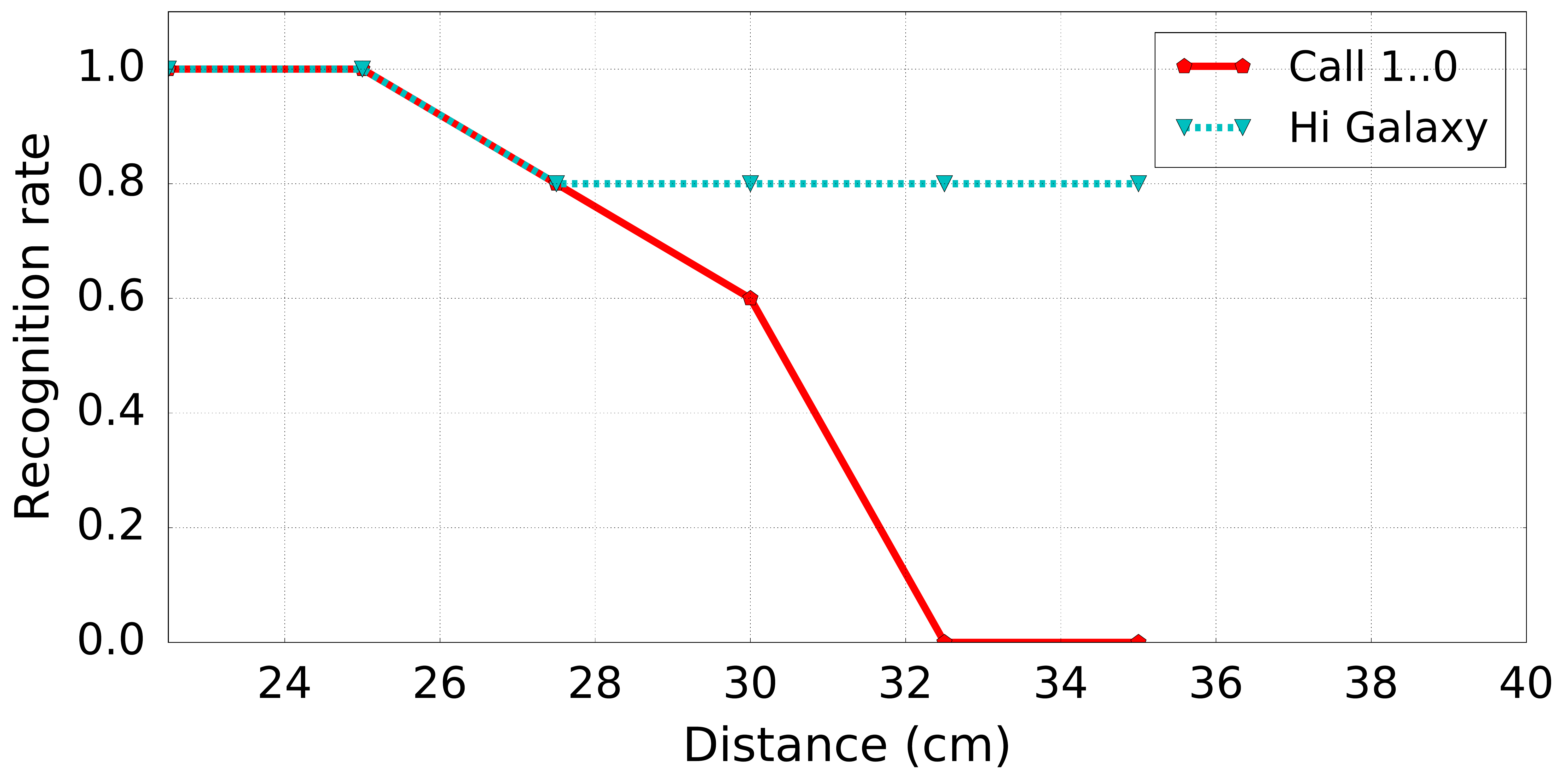}}	
	\subfigure[The recognition rates of the Apple watch]{
		\includegraphics[width=0.46\textwidth]{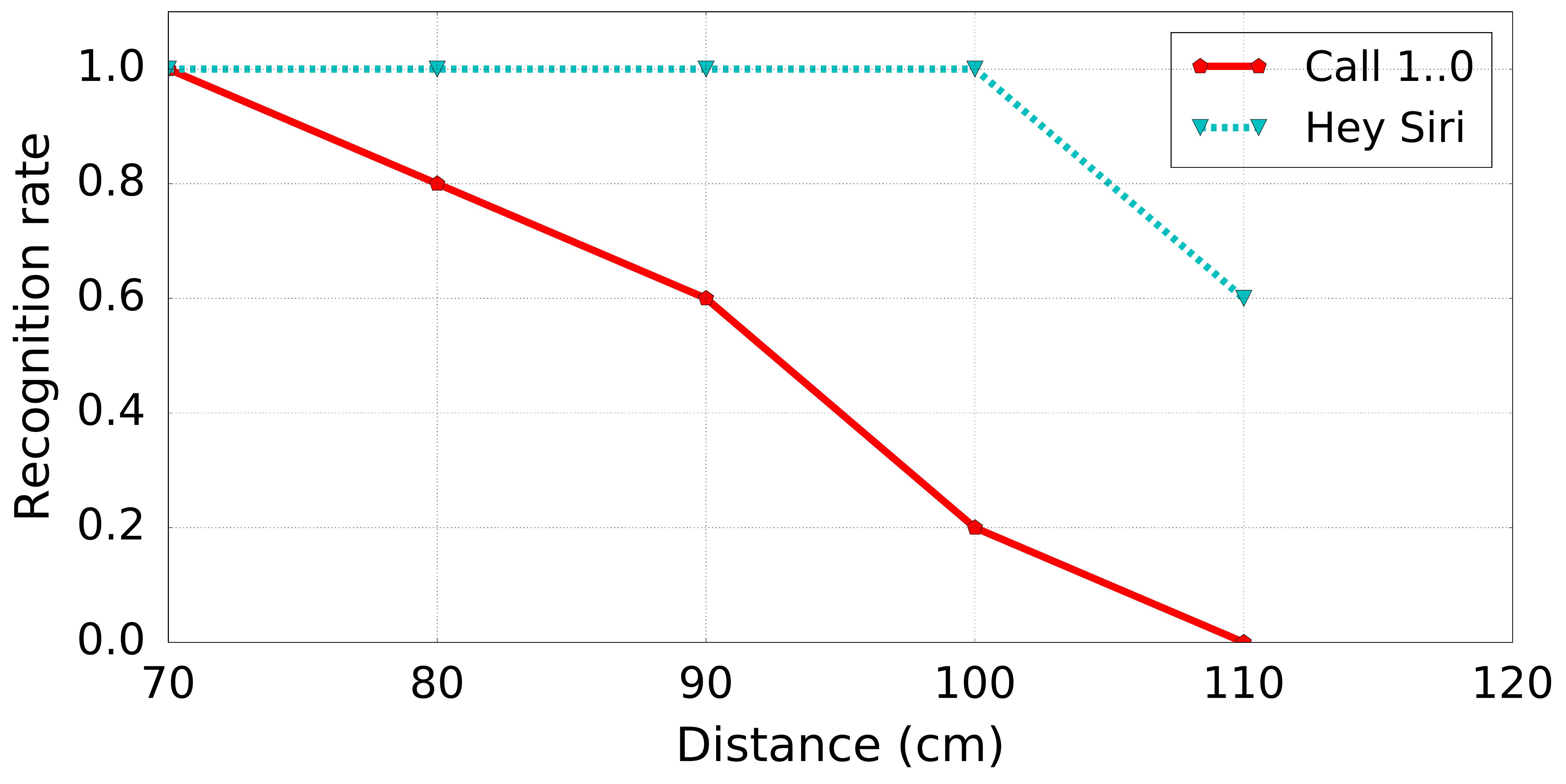}}
	%\vspace{-0.1in}
	\caption{The impact of attack distances on the recognition rates for two portable devices.}
	\label{Fig:Distance}
	%\vspace{-0.1in}
\end{figure*}

\subsection{Impact of Background Noises}
Speech recognition is known to be sensitive to background noises and is recommended to be used in a quiet environment.  %Being inaudible, we envision that \doa can inject an inaudible voice command at a high SPL such that the noise to signal ratio of the voice commands is guaranteed even if the environment is noisy.  
Thus, we examine inaudible voice command injection via \doa in three scenarios: at an office, in a cafe, and on the street. To ensure that the experiment can be repeatable, we simulate the three scenarios by playing background sounds at a chosen SPL and evaluate their impact on the recognition rates. We choose an Apple watch as the attack target, and measure the background noise by a mini sound meter.

From Tab.~\ref{tab:scene}, we can see that recognition rates for activation command are over 90\% for all the three scenes while the recognition rates of the control command (``Turn on airplane mode'') decrease with the increase of ambient noise levels. That is because the activation command is shorter than the control command. With the increase of the word count for a control command, the recognition rate drops quickly because failure to recognize any word could render the command recognition unsuccessful. 

\subsection{Impact of Sound Pressure Levels}
\label{sec:SPL}

For both audible and inaudible sounds, a higher SPL leads to a better quality of recorded voices and thus a higher recognition rate. This is because a higher SPL always means a larger signal-to-noise ratio (SNR) for given noise levels. To explore the impact of SPLs on \doa, we test the control command (``Call 1234567890'') on both the Apple watch and a Galaxy S6 Edge smartphone. In all experiments, the speaker is placed 10 cm from the target device, and the mini sound meter is placed alongside the speaker to measure the environment noise.

We quantify the impact of SPLs with two granularities: sentence recognition rates and word recognition rates. Sentence recognition rates calculate the percentage of successfully recognized commands. Only if every word in the command is recognized correctly, the command is considered to be recognized. % to represents the percentage of voice commands that can be fully recognized and correctly executed by the VCS. However, we find that even one word in the sentence is recognized wrongly, the command won't be correctly executed by the VCS, meaning that sentence rate is not very sensitive to changing SPL. Thus, we come up with a 
Word recognition rates are the percentage of words that are correctly interpreted. For example, if the command ``call 1234567890'' is recognized as ``call 1234567'', the words recognition rate is 63.6\% (7/11). 

Fig.~\ref{Fig:SPL} (a) (b) show the impact of the SPLs on both types of recognition rates. Not surprisingly, %According to the definition of sentence recognition rate and words recognition rate, it is not surprising that 
given the same SPL, the word recognition rates are always larger than the sentence recognition rates until both reach 100\%. For the Apple watch, both recognition rates become 100\% once the SPL is larger than 106.2 dB. %the words and sentence recognition rate is increasing flatly with the growth of SPL. When , the success rates of sentence and words are 100 \%. The performance of Galaxy in sentence recognition was not as good as words recognition. And we can also find that 
In comparison, the minimum SPL for the Galaxy S6 Edge to achieve a 100\% recognition rate is 113.96 dB, which is higher than that of the Apple watch. This is because the Apple watch outperforms the Galaxy S6 Edge in terms of demodulating inaudible voice commands. % has better response to high frequency signals. 

\subsection{Impact of Attack Distances}

In this section, an activation command (either ``Hey Siri'' or ``Hi Galaxy'') and a control command (``Call 1234567890'') are used to test the recognition rates at various distances. We evaluate the recognition rates of two commands on an Apple watch and a Galaxy S6 Edge, and we depict the results in Fig.~\ref{Fig:Distance}.

In general, the recognition rates of the activation command are  higher than that of the control command, because the activation command contains a smaller number of words than the control command. %it is consistent with the previous results in section ~\ref{sec:SPL} and the ideal attack distance are below 70 cm for Apple Watch, 25 cm  for S6 Edge respectively,
The Apple watch can be activated with a success rate of 100\% from 100 cm away, and the Galaxy S6 Edge can be activated with 100\% from 25 cm. We believe that the difference between the two devices is because Apple watches are worn on the wrist and are designed to accept voice commands from a longer distance than a smartphone.  

\subsection{Evaluation of Portable Device Attacks}
In this section, we evaluate the effectiveness of portable device attacks.

\textbf{Setup}. We use the Galaxy S6 Edge smartphone running Android 6.0.1 as the attack device and an Apple watch as the victim device which is paired with an iPhone 6 Plus. The attack voice command is ``turn on airplane mode''. We set $f_c$ to be \{20, 21, 22, 23, 24\}~kHz, respectively. The AM depth is 100\%, and the sampling rate is 192~kHz. The baseband signal has a maximum frequency of 3~kHz.

\textbf{Results}. As shown in Tab.~\ref{tag: S6}, we successfully ``turned on airplane mode'' on the Apple watch at the 23~kHz carrier frequency. Note that 20~kHz and 21~kHz are also successful. However, there are frequency leakages below 20~kHz and it sounds like crickets and can be heard. The word and sentence recognition rates are 100\%. With the increase of $f_c$, the Apple watch fails to recognize the voice command because of frequency selectivity of the speaker.
\begin{table}[tb]
	\centering
	\caption{Portable device attack results. Attacking an Apple watch using a Galaxy S6 Edge smartphone that is 2~cm away.} %\normalfont{We can achieve a 100\% success rate at an attack distance smaller than 2 cm.}}
	\label{tag: S6}
	\begin{threeparttable}
		\begin{tabular}{c|c|c|c|c|c}
			\hline
			\textbf{$f_c$ (kHz)} & \textbf{20} & \textbf{21} & \textbf{22} & \textbf{23} & \textbf{24}\\
			\hline
			\hline
			Word recognition rate & 80\% &  100\% & 16\% & 100\% & 0\%\\
			\hline
			Sentence recognition rate & 80\% & 100\% & 0\% & 100\% & 0\%\\
			\hline
		\end{tabular}
	\end{threeparttable}
	%\vspace{-5pt}
\end{table}

To extend the attack distance, we utilize a low-power audio amplifier (3 Watt) module to drive an ultrasonic transducer, as is shown in Fig.~\ref{fig:transmitter}. 
%We use Galaxy S6 Edge instead of iPhone SE as our attack device, because iPhone SE audio interface couldn't emit out signal above 24 kHz. The Galaxy S6 Edge offers audio playback up to 24-bit/192 kHz with a wide range of formats accepted including FLAC and WAV. 
With the amplifier module, the maximum distance of effective attacks is increased to 27 cm. %In our evaluation, we only implemented a simple amplifier circuit. 
Note that the attack distance can be further extended with professional devices and more powerful amplifiers.

The adversary can launch a remote attack utilizing a victim's device. For example, an adversary can upload an audio or video clip in which the voice commands are embedded in a website, e.g. YouTube. When the audio or video is played by the victims' devices, the surrounding voice controllable systems such as Google Home assistant, Alexa, and mobile phones may be triggered unconsciously.

\section{Defenses}
\label{sec:defense}
In this section, we discuss the defense strategies to address the aforementioned attacks from both the hardware and software perspectives.  % to SR systems revealed 

\subsection{Hardware-Based Defense} 

We propose two hardware-based defense strategies: microphone enhancement and baseband cancellation. 

\textbf{Microphone Enhancement.} The root cause of inaudible voice commands is that microphones can sense acoustic sounds with a frequency higher than 20 kHz while an ideal microphone should not. By default, most MEMS microphones on mobile devices nowadays allow signals above 20 kHz~\cite{Akustica1, Akustica2, STMicroelectronics1, STMicroelectronics2, Knowles1}. Thus, a microphone shall be enhanced and designed to suppress any acoustic signals whose frequencies are in the ultrasound range. For instance, the microphone of iPhone 6 Plus can resist to inaudible voice commands well.

\textbf{Inaudible Voice Command Cancellation.} Given the legacy microphones, we can add a module prior to LPF to detect the modulated voice commands and cancel the baseband with the modulated voice commands. In particular, we can detect the signals within the ultrasound frequency range that exhibit AM modulation characteristics, and demodulate the signals to obtain the baseband. For instance, in the presence of inaudible voice command injection, besides the demodulated baseband signals $m(t)$, the recorded analog voice signals shall include the original modulated signal:
$	v(t) = A m(t)\cos(2\pi f_ct) + \cos(2\pi f_ct),$
where $A$ is the gain for the input signal $m(t)$. 
By down-converting $v(t)$ to obtain $Am(t)$ and adjusting the amplitude, we can subtract the baseband signal. Note that such a command cancellation procedure will not affect the normal operation of a microphone, since there will be no correlation between the captured audible voice signals and noises in the ultrasound range.

\subsection{Software-Based Defense}
Software-based defense looks into the unique features of modulated voice commands which are distinctive from genuine ones.

As shown in Fig.~\ref{Fig:recoveredsound}, the recovered (demodulated) attack signal shows differences from both the original signal  and the recorded one in the high frequency ranging from 500 to 1000~Hz. The original signal is produced by the Google TTS engine, the carrier frequency for modulation is 25~kHz. Thus, we can detect \doa by analyzing the signal in the frequency range from 500 to 1000~Hz. In particular, a machine learning based classifier shall detect it.

\begin{figure}[tp]
	\includegraphics[width=0.496\textwidth]{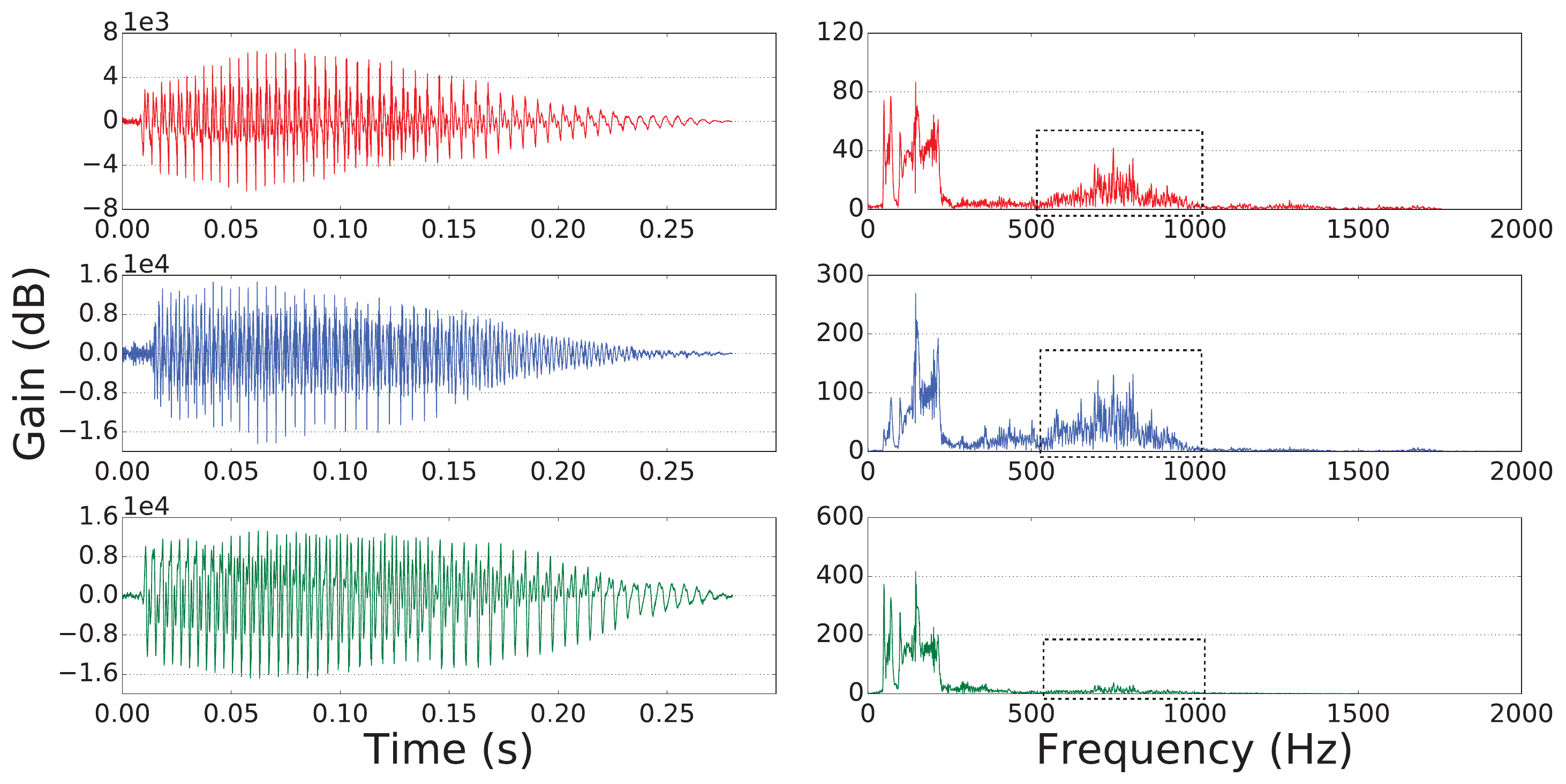}
%	\captionsetup{belowskip=-10pt}
	\caption{Original (top), recorded (middle) and recovered (bottom) voice signals. The modulated voice command differs from both the original signal and the recorded one in the frequency range between 500 and 1000 Hz.} 
	\label{Fig:recoveredsound}	
%	\vspace{-0.2in}
\end{figure}

To validate the feasibility of detecting \doa, we utilize supported vector machine (SVM) as the classifier, and  extracted 15 features in the time and frequency domains from audios. We generated 12 voice commands (i.e., ``Hey Siri''): 8 types of voices from the NeoSpeech TTS engine and 4 types of voices from the Selvy TTS engine. With each type, we obtained two samples: one is recorded and the other is recovered. In total, we have 24 samples. To train a SVM classifier, we use 5 recorded audios as positive samples and 5 recovered audios as negative samples. The rest 14 samples are used for testing. The classifier can distinguish the recovered audios from recorded ones with 100\% true positive rate (7/7) and 100\% true negative rate (7/7). The result using a simple SVM classifier indicating that software-based defense strategy can be used to detect \doa.

% !TEX root = CCS2017_DolphinAttack.tex
\section{Related Work}
\label{sec:relatedWork}
%Related work of \doa can be grouped into three categories.
%\yc{Reviewer1: can be expanded regarding the use of acoustic channel in other applications}

\textbf{Security of voice controllable systems.} An increasing amount of research effort is devoted into studying the security of voice controllable systems~\cite{kasmi2015iemi,mukhopadhyay2015all,diao2014your,vaidya2015cocaine, carlini2016hidden}. Kasmi et al. ~\cite{kasmi2015iemi} introduced a voice command injection attack against modern smartphones by applying intentional electromagnetic interference on headphone cables, while in this paper, we inject voice commands by utilizing the nonlinearity of microphones over ultrasounds.
% The limitation is that the attacked devices must be plugged in a smartphone. 
Mukhopadhyay et al.~\cite{mukhopadhyay2015all} demonstrated voice impersonation attacks on state-of-the-art automated speaker verification algorithms. They built a model of the victim's voice based on the samples from the victim. Diao et al. ~\cite{diao2014your} designed permission bypass attacks from a zero-permission Android application through phone speakers. 
% This attack requires a malicious app to be installed on the target device, and the voice commands played by speaker can be heard by the victim.
Hidden voice commands and Cocaine noodles~\cite{vaidya2015cocaine, carlini2016hidden} use audible and mangled audio commands to attack speech recognition systems. Under these attacks, the victims can observe the obfuscated voice commands sometimes. \doa is motivated by these attacks, but is completely inaudible and imperceptible and we show it is possible to launch \doa using portable devices.

\textbf{Security of sensor-equipped devices.} Commercial devices equipped with various sensors (e.g., smartphones, wearables, and tablets) are gaining their popularity. Along with the growing trend of ubiquitous mobile devices are the security concerns.
Many researchers~\cite{shin2016sampling,dean2007degradation,dean2011characterization,trippel2017walnut} focus on studying possible attacks against sensors on smart devices. Among which, sensor spoofing (i.e., the injection of a malicious signal into a victim sensor) has attracted much attention  and is considered one of the most critical threats to sensor-equipped devices. Shin et al. investigate and classify sensor spoofing attacks~\cite{shin2016sampling} into three categories: regular channel attacks (replay attack)~\cite{francillon2011relay,yan2016can,ishtiaq2010security}, transmission channel attacks, and side channel attacks~\cite{dean2007degradation, dean2011characterization,son2015rocking,castro2007influence, shoukry2013non}. Dean et al.~\cite{dean2007degradation, dean2011characterization} demonstrated that MEMS gyroscopes are susceptible to high-power high-frequency acoustic noises when acoustic frequency components are close to the resonating frequency of the gyroscope's sensing mass. Utilizing on-board sensors, Gu et al. \cite{gu2014toauth} designed a cryptographic key generation mechanism by using vibration motors and accelerometers. Our work focuses on microphones, which can be considered as one type of sensors.

\textbf{Privacy leakage through sensors.} Michalevsky et al.~\cite{Yan2014Gyrophone} utilized MEMS gyroscopes to measure acoustic signals, which reveal the speaker information. Schlegel et al.~\cite{Schlegel2011Soundcomber} designed a Trojan that can extract high-value data from audio sensors of smartphones. Owusu et al.~\cite{Owusu2006ACCessory} utilized the accelerometer readings as a side channel to extract the entire sequences of entered text on a smartphone touchscreen keyboard without requiring special privileges. Aviv et al. \cite{Aviv2012Practicality} demonstrated that accelerometer sensors can reveal user taps and gesture-based input. Dey et al.~\cite{dey2014accelprint} studied how to fingerprint smartphones utilizing the imperfections of on-board accelerometers, and the fingerprints can act as an identifier to track the smartphone's owner.
Simon et al.~\cite{Simon2013PIN} utilized video cameras and microphones to infer PINs entered on a number-only soft keyboard on a smartphone. Li et al. \cite{li2017you} can verify the capture time and location of the photos with the sun position estimated based on the shadows in the photo and sensor readings of the cameras.
Sun et al.~\cite{sun2016visible} presented a video-assisted keystroke inference framework to infer a tablet user's inputs from surreptitious video recordings of the tablet motion. Backes et al.~\cite{backes2010acoustic} showed it is possible to recover what a dot matrix printer processing English text is printing based on the printer's acoustic noises. Similarly, we study how to utilize microphone vulnerabilities for security and privacy breaches.

Roy et al.~\cite{roy2017backdoor} presented BackDoor, which constructs an acoustic (but inaudible) communication channel between two speakers and a microphone over ultrasound bands. In particular, BackDoor utilizes two ultrasonic speakers to transmit two frequencies. After passing through the microphone's non-linear diaphragm and power-amplifier, the two signals create a ``shadow'' in the audible frequency range, which could carry data. 
However, the ``shadow'' is a single tone instead of a voice command that consists of a rich set of tones. In comparison, we show it is possible to use one speaker to inject inaudible commands to SR systems, causing various security and privacy issues.

% !TEX root = CCS2017_DolphinAttack.tex
\section{Conclusion}
\label{sec:conclusion}
In this paper, we propose \doa, an inaudible attack to SR systems. \doa leverages the AM (amplitude modulation) technique to modulate audible voice commands on ultrasonic carriers by which the command signals can not be perceived by human. With \doa, an adversary can attack major SR systems including Siri, Google Now, Alexa, and etc. To avoid the abuse of \doa in reality, we propose two defense solutions from the aspects of both hardware and software.

\begin{acks}

This work has been funded in part by 360 Technology Inc., NSFC 61472358, NSFC 61702451, NSF CNS-0845671, and the Fundamental Research Funds for the Central Universities 2017QNA4017.
\end{acks}

\bibliographystyle{ACM-Reference-Format}
\bibliography{reference}


\begin{thebibliography}{00}

%%% the Web address.

\ifx \showCODEN    \undefined \def \showCODEN     #1{\unskip}     \fi
\ifx \showDOI      \undefined \def \showDOI       #1{#1}\fi
\ifx \showISBNx    \undefined \def \showISBNx     #1{\unskip}     \fi
\ifx \showISBNxiii \undefined \def \showISBNxiii  #1{\unskip}     \fi
\ifx \showISSN     \undefined \def \showISSN      #1{\unskip}     \fi
\ifx \showLCCN     \undefined \def \showLCCN      #1{\unskip}     \fi
\ifx \shownote     \undefined \def \shownote      #1{#1}          \fi
\ifx \showarticletitle \undefined \def \showarticletitle #1{#1}   \fi
\ifx \showURL      \undefined \def \showURL       {\relax}        \fi
% invisible to TeX
\providecommand\bibfield[2]{#2}
\providecommand\bibinfo[2]{#2}
\providecommand\natexlab[1]{#1}
\providecommand\showeprint[2][]{arXiv:#2}

\bibitem[\protect\citeauthoryear{Abuelma'atti}{Abuelma'atti}{2003}]%
        {abuelma2003analysis}
\bibfield{author}{\bibinfo{person}{Muhammad~Taher Abuelma'atti}.}
  \bibinfo{year}{2003}\natexlab{}.
\newblock \showarticletitle{Analysis of the effect of radio frequency
  interference on the DC performance of bipolar operational amplifiers}.
\newblock \bibinfo{journal}{{\em IEEE Transactions on Electromagnetic
  Compatibility\/}} \bibinfo{volume}{45}, \bibinfo{number}{2}
  (\bibinfo{year}{2003}), \bibinfo{pages}{453--458}.
\newblock


\bibitem[\protect\citeauthoryear{Akustica}{Akustica}{2014a}]%
        {Akustica2}
\bibfield{author}{\bibinfo{person}{Akustica}.}
  \bibinfo{year}{2014}\natexlab{a}.
\newblock \bibinfo{title}{{AKU143 Top Port, analog silicon MEMS microphone}}.
\newblock
  \bibinfo{howpublished}{\url{http://www.mouser.com/ds/2/720/DS37-1.01\%20AKU143\%20Datasheet-552974.pdf}}.
    (\bibinfo{year}{2014}).
\newblock


\bibitem[\protect\citeauthoryear{Akustica}{Akustica}{2014b}]%
        {Akustica1}
\bibfield{author}{\bibinfo{person}{Akustica}.}
  \bibinfo{year}{2014}\natexlab{b}.
\newblock \bibinfo{title}{{AKU242 digital silicon MEMS microphone}}.
\newblock
  \bibinfo{howpublished}{\url{http://www.mouser.com/ds/2/720/PB24-1.0\%20-\%20AKU242\%20Product\%20Brief-770082.pdf}}.
    (\bibinfo{year}{2014}).
\newblock


\bibitem[\protect\citeauthoryear{Amazon}{Amazon}{2017}]%
        {Alexa}
\bibfield{author}{\bibinfo{person}{Amazon}.} \bibinfo{year}{2017}\natexlab{}.
\newblock \bibinfo{title}{{Alexa}}.
\newblock \bibinfo{howpublished}{\url{https://developer.amazon.com/alexa}}.
  (\bibinfo{year}{2017}).
\newblock


\bibitem[\protect\citeauthoryear{Apple}{Apple}{2017}]%
        {Siri}
\bibfield{author}{\bibinfo{person}{Apple}.} \bibinfo{year}{2017}\natexlab{}.
\newblock \bibinfo{title}{{iOS-Siri-Apple}}.
\newblock \bibinfo{howpublished}{\url{https://www.apple.com/ios/siri/}}.
  (\bibinfo{year}{2017}).
\newblock


\bibitem[\protect\citeauthoryear{Aviv, Sapp, Blaze, and Smith}{Aviv
  et~al\mbox{.}}{2012}]%
        {Aviv2012Practicality}
\bibfield{author}{\bibinfo{person}{Adam~J. Aviv}, \bibinfo{person}{Benjamin
  Sapp}, \bibinfo{person}{Matt Blaze}, {and} \bibinfo{person}{Jonathan~M.
  Smith}.} \bibinfo{year}{2012}\natexlab{}.
\newblock \showarticletitle{Practicality of accelerometer side channels on
  smartphones}. In \bibinfo{booktitle}{{\em Proceedings of the Computer
  Security Applications Conference}}. \bibinfo{pages}{41--50}.
\newblock


\bibitem[\protect\citeauthoryear{Backes, D{\"u}rmuth, Gerling, Pinkal, and
  Sporleder}{Backes et~al\mbox{.}}{2010}]%
        {backes2010acoustic}
\bibfield{author}{\bibinfo{person}{Michael Backes}, \bibinfo{person}{Markus
  D{\"u}rmuth}, \bibinfo{person}{Sebastian Gerling}, \bibinfo{person}{Manfred
  Pinkal}, {and} \bibinfo{person}{Caroline Sporleder}.}
  \bibinfo{year}{2010}\natexlab{}.
\newblock \showarticletitle{Acoustic side-channel attacks on printers.}. In
  \bibinfo{booktitle}{{\em Proceedings of the USENIX Security Symposium}}.
  \bibinfo{pages}{307--322}.
\newblock


\bibitem[\protect\citeauthoryear{Baidu}{Baidu}{2017}]%
        {Baidu}
\bibfield{author}{\bibinfo{person}{Baidu}.} \bibinfo{year}{2017}\natexlab{}.
\newblock \bibinfo{title}{{Baidu Translate}}.
\newblock \bibinfo{howpublished}{\url{http://fanyi.baidu.com/}}.
  (\bibinfo{year}{2017}).
\newblock


\bibitem[\protect\citeauthoryear{Bioacoustics}{Bioacoustics}{2017}]%
        {vifa}
\bibfield{author}{\bibinfo{person}{Avisoft Bioacoustics}.}
  \bibinfo{year}{2017}\natexlab{}.
\newblock \bibinfo{title}{{Ultrasonic Dynamic Speaker Vifa}}.
\newblock \bibinfo{howpublished}{\url{http://www.avisoft.com/usg/vifa.htm}}.
  (\bibinfo{year}{2017}).
\newblock


\bibitem[\protect\citeauthoryear{Carlini, Mishra, Vaidya, Zhang, Sherr,
  Shields, Wagner, and Zhou}{Carlini et~al\mbox{.}}{2016}]%
        {carlini2016hidden}
\bibfield{author}{\bibinfo{person}{Nicholas Carlini}, \bibinfo{person}{Pratyush
  Mishra}, \bibinfo{person}{Tavish Vaidya}, \bibinfo{person}{Yuankai Zhang},
  \bibinfo{person}{Micah Sherr}, \bibinfo{person}{Clay Shields},
  \bibinfo{person}{David Wagner}, {and} \bibinfo{person}{Wenchao Zhou}.}
  \bibinfo{year}{2016}\natexlab{}.
\newblock \showarticletitle{Hidden voice commands}. In \bibinfo{booktitle}{{\em
  Proceedings of the USENIX Security Symposium}}.
\newblock


\bibitem[\protect\citeauthoryear{Castro, Dean, Roth, Flowers, and
  Grantham}{Castro et~al\mbox{.}}{2007}]%
        {castro2007influence}
\bibfield{author}{\bibinfo{person}{Simon Castro}, \bibinfo{person}{Robert
  Dean}, \bibinfo{person}{Grant Roth}, \bibinfo{person}{George~T Flowers},
  {and} \bibinfo{person}{Brian Grantham}.} \bibinfo{year}{2007}\natexlab{}.
\newblock \showarticletitle{Influence of acoustic noise on the dynamic
  performance of MEMS gyroscopes}. In \bibinfo{booktitle}{{\em Proceedings of
  the ASME International Mechanical Engineering Congress and Exposition}}.
  American Society of Mechanical Engineers, \bibinfo{pages}{1825--1831}.
\newblock


\bibitem[\protect\citeauthoryear{CereProc}{CereProc}{2017}]%
        {cereproc}
\bibfield{author}{\bibinfo{person}{CereProc}.} \bibinfo{year}{2017}\natexlab{}.
\newblock \bibinfo{title}{{CereProc Text-to-Speech}}.
\newblock \bibinfo{howpublished}{\url{https://www.cereproc.com/}}.
  (\bibinfo{year}{2017}).
\newblock


\bibitem[\protect\citeauthoryear{Chen and Whalen}{Chen and Whalen}{1981}]%
        {chen1981comparative}
\bibfield{author}{\bibinfo{person}{Gordon~KC Chen} {and}
  \bibinfo{person}{James~J Whalen}.} \bibinfo{year}{1981}\natexlab{}.
\newblock \showarticletitle{Comparative RFI performance of bipolar operational
  amplifiers}. In \bibinfo{booktitle}{{\em Proceedings of the IEEE
  International Symposium on Electromagnetic Compatibility}}. IEEE,
  \bibinfo{pages}{1--5}.
\newblock


\bibitem[\protect\citeauthoryear{Dean, Castro, Flowers, Roth, Ahmed, Hodel,
  Grantham, Bittle, and Brunsch}{Dean et~al\mbox{.}}{2011}]%
        {dean2011characterization}
\bibfield{author}{\bibinfo{person}{Robert~Neal Dean},
  \bibinfo{person}{Simon~Thomas Castro}, \bibinfo{person}{George~T Flowers},
  \bibinfo{person}{Grant Roth}, \bibinfo{person}{Anwar Ahmed},
  \bibinfo{person}{Alan~Scottedward Hodel}, \bibinfo{person}{Brian~Eugene
  Grantham}, \bibinfo{person}{David~Allen Bittle}, {and}
  \bibinfo{person}{James~P Brunsch}.} \bibinfo{year}{2011}\natexlab{}.
\newblock \showarticletitle{A characterization of the performance of a MEMS
  gyroscope in acoustically harsh environments}.
\newblock \bibinfo{journal}{{\em IEEE Transactions on Industrial
  Electronics\/}} \bibinfo{volume}{58}, \bibinfo{number}{7}
  (\bibinfo{year}{2011}), \bibinfo{pages}{2591--2596}.
\newblock


\bibitem[\protect\citeauthoryear{Dean, Flowers, Hodel, Roth, Castro, Zhou,
  Moreira, Ahmed, Rifki, Grantham, et~al\mbox{.}}{Dean et~al\mbox{.}}{2007}]%
        {dean2007degradation}
\bibfield{author}{\bibinfo{person}{Robert~N Dean}, \bibinfo{person}{George~T
  Flowers}, \bibinfo{person}{A~Scotte Hodel}, \bibinfo{person}{Grant Roth},
  \bibinfo{person}{Simon Castro}, \bibinfo{person}{Ran Zhou},
  \bibinfo{person}{Alfonso Moreira}, \bibinfo{person}{Anwar Ahmed},
  \bibinfo{person}{Rifki Rifki}, \bibinfo{person}{Brian~E Grantham},
  {et~al\mbox{.}}} \bibinfo{year}{2007}\natexlab{}.
\newblock \showarticletitle{On the degradation of MEMS gyroscope performance in
  the presence of high power acoustic noise}. In \bibinfo{booktitle}{{\em
  Proceedings of the IEEE International Symposium on Industrial Electronics}}.
  \bibinfo{pages}{1435--1440}.
\newblock


\bibitem[\protect\citeauthoryear{Devices}{Devices}{2011}]%
        {ADMP401}
\bibfield{author}{\bibinfo{person}{Analog Devices}.}
  \bibinfo{year}{2011}\natexlab{}.
\newblock \bibinfo{title}{ADMP401: Omnidirectional microphone with bottom port
  and analog output obsolete data sheet}.
\newblock
  \bibinfo{howpublished}{\url{http://www.analog.com/media/en/technical-documentation/obsolete-data-sheets/ADMP401.pdf}}.
    (\bibinfo{year}{2011}).
\newblock


\bibitem[\protect\citeauthoryear{Dey, Roy, Xu, Choudhury, and Nelakuditi}{Dey
  et~al\mbox{.}}{2014}]%
        {dey2014accelprint}
\bibfield{author}{\bibinfo{person}{Sanorita Dey}, \bibinfo{person}{Nirupam
  Roy}, \bibinfo{person}{Wenyuan Xu}, \bibinfo{person}{Romit~Roy Choudhury},
  {and} \bibinfo{person}{Srihari Nelakuditi}.} \bibinfo{year}{2014}\natexlab{}.
\newblock \showarticletitle{AccelPrint: Imperfections of Accelerometers Make
  Smartphones Trackable.}. In \bibinfo{booktitle}{{\em Proceedings of the
  Network and Distributed System Security Symposium (NDSS)}}.
\newblock


\bibitem[\protect\citeauthoryear{Diao, Liu, Zhou, and Zhang}{Diao
  et~al\mbox{.}}{2014}]%
        {diao2014your}
\bibfield{author}{\bibinfo{person}{Wenrui Diao}, \bibinfo{person}{Xiangyu Liu},
  \bibinfo{person}{Zhe Zhou}, {and} \bibinfo{person}{Kehuan Zhang}.}
  \bibinfo{year}{2014}\natexlab{}.
\newblock \showarticletitle{Your voice assistant is mine: How to abuse speakers
  to steal information and control your phone}. In \bibinfo{booktitle}{{\em
  Proceedings of the ACM Workshop on Security and Privacy in Smartphones \&
  Mobile Devices}}. ACM, \bibinfo{pages}{63--74}.
\newblock


\bibitem[\protect\citeauthoryear{Francillon, Danev, and Capkun}{Francillon
  et~al\mbox{.}}{2011}]%
        {francillon2011relay}
\bibfield{author}{\bibinfo{person}{Aur{\'e}lien Francillon},
  \bibinfo{person}{Boris Danev}, {and} \bibinfo{person}{Srdjan Capkun}.}
  \bibinfo{year}{2011}\natexlab{}.
\newblock \showarticletitle{Relay attacks on passive keyless entry and start
  systems in modern cars}. In \bibinfo{booktitle}{{\em Proceedings of the
  Network and Distributed System Security Symposium (NDSS)}}.
\newblock


\bibitem[\protect\citeauthoryear{Gago, Balcells, Gonz{\'A}lez, Lamich, Mon, and
  Santolaria}{Gago et~al\mbox{.}}{2007}]%
        {gago2007emi}
\bibfield{author}{\bibinfo{person}{Javier Gago}, \bibinfo{person}{Josep
  Balcells}, \bibinfo{person}{David Gonz{\'A}lez}, \bibinfo{person}{Manuel
  Lamich}, \bibinfo{person}{Juan Mon}, {and} \bibinfo{person}{Alfonso
  Santolaria}.} \bibinfo{year}{2007}\natexlab{}.
\newblock \showarticletitle{EMI susceptibility model of signal conditioning
  circuits based on operational amplifiers}.
\newblock \bibinfo{journal}{{\em IEEE Transactions on Electromagnetic
  Compatibility\/}} \bibinfo{volume}{49}, \bibinfo{number}{4}
  (\bibinfo{year}{2007}), \bibinfo{pages}{849--859}.
\newblock


\bibitem[\protect\citeauthoryear{Google}{Google}{2016}]%
        {GoogleNow}
\bibfield{author}{\bibinfo{person}{Google}.} \bibinfo{year}{2016}\natexlab{}.
\newblock \bibinfo{title}{{Google Now}}.
\newblock
  \bibinfo{howpublished}{\url{http://www.androidcentral.com/google-now}}.
  (\bibinfo{year}{2016}).
\newblock


\bibitem[\protect\citeauthoryear{Group}{Group}{2017}]%
        {acapela}
\bibfield{author}{\bibinfo{person}{Acapela Group}.}
  \bibinfo{year}{2017}\natexlab{}.
\newblock \bibinfo{title}{Acapela text to speech demo}.
\newblock \bibinfo{howpublished}{\url{http://www.acapela-group.com/}}.
  (\bibinfo{year}{2017}).
\newblock


\bibitem[\protect\citeauthoryear{Group}{Group}{2012}]%
        {Festvox}
\bibfield{author}{\bibinfo{person}{CMU~Speech Group}.}
  \bibinfo{year}{2012}\natexlab{}.
\newblock \bibinfo{title}{{Statistical parametirc sythesis and voice conversion
  techniques}}.
\newblock
  \bibinfo{howpublished}{\url{http://festvox.org/11752/slides/lecture11a.pdf}}.
    (\bibinfo{year}{2012}).
\newblock


\bibitem[\protect\citeauthoryear{Gu, Yang, Shangguan, Ji, and Zhao}{Gu
  et~al\mbox{.}}{2014}]%
        {gu2014toauth}
\bibfield{author}{\bibinfo{person}{Weixi Gu}, \bibinfo{person}{Zheng Yang},
  \bibinfo{person}{Longfei Shangguan}, \bibinfo{person}{Xiaoyu Ji}, {and}
  \bibinfo{person}{Yiyang Zhao}.} \bibinfo{year}{2014}\natexlab{}.
\newblock \showarticletitle{Toauth: Towards automatic near field authentication
  for smartphones}. In \bibinfo{booktitle}{{\em Proceedings of the IEEE 13th
  International Conference on Trust, Security and Privacy in Computing and
  Communications (TrustCom)}}. IEEE, \bibinfo{pages}{229--236}.
\newblock


\bibitem[\protect\citeauthoryear{Horowitz and Hill}{Horowitz and Hill}{1989}]%
        {mixer}
\bibfield{author}{\bibinfo{person}{Paul Horowitz} {and}
  \bibinfo{person}{Winfield Hill}.} \bibinfo{year}{1989}\natexlab{}.
\newblock \bibinfo{booktitle}{{\em The art of electronics}}.
\newblock \bibinfo{publisher}{Cambridge Univ. Press}.
\newblock


\bibitem[\protect\citeauthoryear{Ishtiaq~Roufa, Mustafaa, Travis~Taylora, Xua,
  Gruteserb, Trappeb, and Seskarb}{Ishtiaq~Roufa et~al\mbox{.}}{2010}]%
        {ishtiaq2010security}
\bibfield{author}{\bibinfo{person}{Rob~Millerb Ishtiaq~Roufa},
  \bibinfo{person}{Hossen Mustafaa}, \bibinfo{person}{Sangho~Ohb
  Travis~Taylora}, \bibinfo{person}{Wenyuan Xua}, \bibinfo{person}{Marco
  Gruteserb}, \bibinfo{person}{Wade Trappeb}, {and} \bibinfo{person}{Ivan
  Seskarb}.} \bibinfo{year}{2010}\natexlab{}.
\newblock \showarticletitle{Security and privacy vulnerabilities of in-car
  wireless networks: A tire pressure monitoring system case study}. In
  \bibinfo{booktitle}{{\em Proceedings of the USENIX Security Symposium}}.
  \bibinfo{pages}{11--13}.
\newblock


\bibitem[\protect\citeauthoryear{Ittichaichareon, Suksri, and
  Yingthawornsuk}{Ittichaichareon et~al\mbox{.}}{2012}]%
        {ittichaichareon2012speech}
\bibfield{author}{\bibinfo{person}{Chadawan Ittichaichareon},
  \bibinfo{person}{Siwat Suksri}, {and} \bibinfo{person}{Thaweesak
  Yingthawornsuk}.} \bibinfo{year}{2012}\natexlab{}.
\newblock \showarticletitle{Speech recognition using MFCC}. In
  \bibinfo{booktitle}{{\em Proceedings of the International Conference on
  Computer Graphics, Simulation and Modeling (ICGSM)}}.
  \bibinfo{pages}{28--29}.
\newblock


\bibitem[\protect\citeauthoryear{Kasmi and Esteves}{Kasmi and Esteves}{2015}]%
        {kasmi2015iemi}
\bibfield{author}{\bibinfo{person}{Chaouki Kasmi} {and}
  \bibinfo{person}{Jose~Lopes Esteves}.} \bibinfo{year}{2015}\natexlab{}.
\newblock \showarticletitle{IEMI threats for information security: Remote
  command injection on modern smartphones}.
\newblock \bibinfo{journal}{{\em IEEE Transactions on Electromagnetic
  Compatibility\/}} \bibinfo{volume}{57}, \bibinfo{number}{6}
  (\bibinfo{year}{2015}), \bibinfo{pages}{1752--1755}.
\newblock


\bibitem[\protect\citeauthoryear{Knowles}{Knowles}{2013}]%
        {Knowles1}
\bibfield{author}{\bibinfo{person}{Knowles}.} \bibinfo{year}{2013}\natexlab{}.
\newblock \bibinfo{title}{{SPU0410LR5H-QB Zero-Height SiSonicTM Microphone}}.
\newblock
  \bibinfo{howpublished}{\url{http://www.mouser.com/ds/2/218/-532675.pdf}}.
  (\bibinfo{year}{2013}).
\newblock


\bibitem[\protect\citeauthoryear{Krzywdzinski}{Krzywdzinski}{2017}]%
        {ultrasonicAnalyzer}
\bibfield{author}{\bibinfo{person}{Dexus~Pawel Krzywdzinski}.}
  \bibinfo{year}{2017}\natexlab{}.
\newblock \bibinfo{title}{{Ultrasonic analyzer for iPad and iPhone}}.
\newblock \bibinfo{howpublished}{\url{http://iaudioapps.com/page1/page1.html}}.
    (\bibinfo{year}{2017}).
\newblock


\bibitem[\protect\citeauthoryear{Kune, Backes, Clark, Kramer, Reynolds, Fu,
  Kim, and Xu}{Kune et~al\mbox{.}}{2013}]%
        {kune2013ghost}
\bibfield{author}{\bibinfo{person}{Denis~Foo Kune}, \bibinfo{person}{John
  Backes}, \bibinfo{person}{Shane~S Clark}, \bibinfo{person}{Daniel Kramer},
  \bibinfo{person}{Matthew Reynolds}, \bibinfo{person}{Kevin Fu},
  \bibinfo{person}{Yongdae Kim}, {and} \bibinfo{person}{Wenyuan Xu}.}
  \bibinfo{year}{2013}\natexlab{}.
\newblock \showarticletitle{Ghost talk: Mitigating EMI signal injection attacks
  against analog sensors}. In \bibinfo{booktitle}{{\em Proceedings of the IEEE
  Symposium on Security and Privacy (S\&P)}}. IEEE, \bibinfo{pages}{145--159}.
\newblock


\bibitem[\protect\citeauthoryear{Lee, Kim, Choi, and Choi}{Lee
  et~al\mbox{.}}{2015}]%
        {lee2015chirp}
\bibfield{author}{\bibinfo{person}{Hyewon Lee}, \bibinfo{person}{Tae~Hyun Kim},
  \bibinfo{person}{Jun~Won Choi}, {and} \bibinfo{person}{Sunghyun Choi}.}
  \bibinfo{year}{2015}\natexlab{}.
\newblock \showarticletitle{Chirp signal-based aerial acoustic communication
  for smart devices}. In \bibinfo{booktitle}{{\em Proceedings of the IEEE
  International Conference on Computer Communications (INFOCOM)}}. IEEE,
  \bibinfo{pages}{2407--2415}.
\newblock


\bibitem[\protect\citeauthoryear{Li, Xu, Wang, and Qu}{Li
  et~al\mbox{.}}{2017}]%
        {li2017you}
\bibfield{author}{\bibinfo{person}{Xiaopeng Li}, \bibinfo{person}{Wenyuan Xu},
  \bibinfo{person}{Song Wang}, {and} \bibinfo{person}{Xianshan Qu}.}
  \bibinfo{year}{2017}\natexlab{}.
\newblock \showarticletitle{Are You Lying: Validating the Time-Location of
  Outdoor Images}. In \bibinfo{booktitle}{{\em Proceedings of the International
  Conference on Applied Cryptography and Network Security}}. Springer,
  \bibinfo{pages}{103--123}.
\newblock


\bibitem[\protect\citeauthoryear{Ltd.}{Ltd.}{2017}]%
        {ispectrum}
\bibfield{author}{\bibinfo{person}{Dog Park~Software Ltd.}}
  \bibinfo{year}{2017}\natexlab{}.
\newblock \bibinfo{title}{{iSpectrum - Macintosh Audio Spectrum Analyzer}}.
\newblock
  \bibinfo{howpublished}{\url{https://dogparksoftware.com/iSpectrum.html}}.
  (\bibinfo{year}{2017}).
\newblock


\bibitem[\protect\citeauthoryear{Mateljan}{Mateljan}{2017}]%
        {arta}
\bibfield{author}{\bibinfo{person}{Ivo Mateljan}.}
  \bibinfo{year}{2017}\natexlab{}.
\newblock \bibinfo{title}{Audio measurement and analysis software}.
\newblock \bibinfo{howpublished}{\url{http://www.artalabs.hr/}}.
  (\bibinfo{year}{2017}).
\newblock


\bibitem[\protect\citeauthoryear{Michalevsky, Boneh, and Nakibly}{Michalevsky
  et~al\mbox{.}}{2014}]%
        {Yan2014Gyrophone}
\bibfield{author}{\bibinfo{person}{Yan Michalevsky}, \bibinfo{person}{Dan
  Boneh}, {and} \bibinfo{person}{Gabi Nakibly}.}
  \bibinfo{year}{2014}\natexlab{}.
\newblock \showarticletitle{Gyrophone: Recognizing Speech from Gyroscope
  Signals}. In \bibinfo{booktitle}{{\em Proceedings of the USENIX Security
  Symposium}}. \bibinfo{pages}{1053--1067}.
\newblock


\bibitem[\protect\citeauthoryear{Microsoft}{Microsoft}{2017}]%
        {Cortana}
\bibfield{author}{\bibinfo{person}{Microsoft}.}
  \bibinfo{year}{2017}\natexlab{}.
\newblock \bibinfo{title}{{What is Cortana?}}
\newblock
  \bibinfo{howpublished}{\url{https://support.microsoft.com/en-us/help/17214/windows-10-what-is}}.
    (\bibinfo{year}{2017}).
\newblock


\bibitem[\protect\citeauthoryear{Mukhopadhyay, Shirvanian, and
  Saxena}{Mukhopadhyay et~al\mbox{.}}{2015}]%
        {mukhopadhyay2015all}
\bibfield{author}{\bibinfo{person}{Dibya Mukhopadhyay},
  \bibinfo{person}{Maliheh Shirvanian}, {and} \bibinfo{person}{Nitesh Saxena}.}
  \bibinfo{year}{2015}\natexlab{}.
\newblock \showarticletitle{All your voices are belong to us: Stealing voices
  to fool humans and machines}. In \bibinfo{booktitle}{{\em Proceedings of the
  European Symposium on Research in Computer Security}}. Springer,
  \bibinfo{pages}{599--621}.
\newblock


\bibitem[\protect\citeauthoryear{NeoSpeech}{NeoSpeech}{2017}]%
        {Neospeech}
\bibfield{author}{\bibinfo{person}{NeoSpeech}.}
  \bibinfo{year}{2017}\natexlab{}.
\newblock \bibinfo{title}{{NeoSpeech Text-to-Speech}}.
\newblock \bibinfo{howpublished}{\url{http://www.neospeech.com/}}.
  (\bibinfo{year}{2017}).
\newblock


\bibitem[\protect\citeauthoryear{Owusu, Han, Das, Perrig, and Zhang}{Owusu
  et~al\mbox{.}}{2006}]%
        {Owusu2006ACCessory}
\bibfield{author}{\bibinfo{person}{Emmanuel Owusu}, \bibinfo{person}{Jun Han},
  \bibinfo{person}{Sauvik Das}, \bibinfo{person}{Adrian Perrig}, {and}
  \bibinfo{person}{Joy Zhang}.} \bibinfo{year}{2006}\natexlab{}.
\newblock \showarticletitle{ACCessory: password inference using accelerometers
  on smartphones}.
\newblock  (\bibinfo{year}{2006}).
\newblock


\bibitem[\protect\citeauthoryear{Reinke}{Reinke}{2017}]%
        {spectroid}
\bibfield{author}{\bibinfo{person}{Carl Reinke}.}
  \bibinfo{year}{2017}\natexlab{}.
\newblock \bibinfo{title}{Spectroid}.
\newblock
  \bibinfo{howpublished}{\url{https://play.google.com/store/apps/details?id=org.intoorbit.spectrum&hl=en}}.
    (\bibinfo{year}{2017}).
\newblock


\bibitem[\protect\citeauthoryear{Roy, Hassanieh, and Roy~Choudhury}{Roy
  et~al\mbox{.}}{2017}]%
        {roy2017backdoor}
\bibfield{author}{\bibinfo{person}{Nirupam Roy}, \bibinfo{person}{Haitham
  Hassanieh}, {and} \bibinfo{person}{Romit Roy~Choudhury}.}
  \bibinfo{year}{2017}\natexlab{}.
\newblock \showarticletitle{Backdoor: Making microphones hear inaudible
  sounds}. In \bibinfo{booktitle}{{\em Proceedings of the 15th Annual
  International Conference on Mobile Systems, Applications, and Services}}.
  ACM, \bibinfo{pages}{2--14}.
\newblock


\bibitem[\protect\citeauthoryear{Samsung}{Samsung}{2017}]%
        {SVoice}
\bibfield{author}{\bibinfo{person}{Samsung}.} \bibinfo{year}{2017}\natexlab{}.
\newblock \bibinfo{title}{{What is S Voice?}}
\newblock
  \bibinfo{howpublished}{\url{http://www.samsung.com/global/galaxy/what-is/s-voice/}}.
    (\bibinfo{year}{2017}).
\newblock


\bibitem[\protect\citeauthoryear{Schlegel, Zhang, Zhou, Intwala, Kapadia, and
  Wang}{Schlegel et~al\mbox{.}}{2011}]%
        {Schlegel2011Soundcomber}
\bibfield{author}{\bibinfo{person}{Roman Schlegel}, \bibinfo{person}{Kehuan
  Zhang}, \bibinfo{person}{Xiao-yong Zhou}, \bibinfo{person}{Mehool Intwala},
  \bibinfo{person}{Apu Kapadia}, {and} \bibinfo{person}{XiaoFeng Wang}.}
  \bibinfo{year}{2011}\natexlab{}.
\newblock \showarticletitle{Soundcomber: A Stealthy and Context-Aware Sound
  Trojan for Smartphones}. In \bibinfo{booktitle}{{\em Proceedings of the
  Network and Distributed System Security Symposium (NDSS)}},
  Vol.~\bibinfo{volume}{11}. \bibinfo{pages}{17--33}.
\newblock


\bibitem[\protect\citeauthoryear{Sestek}{Sestek}{2017}]%
        {Sestek}
\bibfield{author}{\bibinfo{person}{Sestek}.} \bibinfo{year}{2017}\natexlab{}.
\newblock \bibinfo{title}{{Sestek TTS}}.
\newblock \bibinfo{howpublished}{\url{http://www.sestek.com/}}.
  (\bibinfo{year}{2017}).
\newblock


\bibitem[\protect\citeauthoryear{Shin, Son, Park, Kwon, and Kim}{Shin
  et~al\mbox{.}}{2016}]%
        {shin2016sampling}
\bibfield{author}{\bibinfo{person}{Hocheol Shin}, \bibinfo{person}{Yunmok Son},
  \bibinfo{person}{Youngseok Park}, \bibinfo{person}{Yujin Kwon}, {and}
  \bibinfo{person}{Yongdae Kim}.} \bibinfo{year}{2016}\natexlab{}.
\newblock \showarticletitle{Sampling race: bypassing timing-based analog active
  sensor spoofing detection on analog-digital systems}. In
  \bibinfo{booktitle}{{\em Proceedings of the USENIX Workshop on Offensive
  Technologies (WOOT)}}. USENIX Association.
\newblock


\bibitem[\protect\citeauthoryear{Shoukry, Martin, Tabuada, and
  Srivastava}{Shoukry et~al\mbox{.}}{2013}]%
        {shoukry2013non}
\bibfield{author}{\bibinfo{person}{Yasser Shoukry}, \bibinfo{person}{Paul
  Martin}, \bibinfo{person}{Paulo Tabuada}, {and} \bibinfo{person}{Mani
  Srivastava}.} \bibinfo{year}{2013}\natexlab{}.
\newblock \showarticletitle{Non-invasive spoofing attacks for anti-lock braking
  systems}. In \bibinfo{booktitle}{{\em Proceedings of the International
  Workshop on Cryptographic Hardware and Embedded Systems}}. Springer,
  \bibinfo{pages}{55--72}.
\newblock


\bibitem[\protect\citeauthoryear{Simon and Anderson}{Simon and
  Anderson}{2013}]%
        {Simon2013PIN}
\bibfield{author}{\bibinfo{person}{Laurent Simon} {and} \bibinfo{person}{Ross
  Anderson}.} \bibinfo{year}{2013}\natexlab{}.
\newblock \showarticletitle{PIN skimmer: inferring PINs through the camera and
  microphone}. In \bibinfo{booktitle}{{\em Proceedings of the ACM Workshop on
  Security and Privacy in Smartphones \& Mobile Devices}}.
  \bibinfo{pages}{67--78}.
\newblock


\bibitem[\protect\citeauthoryear{Son, Shin, Kim, Park, Noh, Choi, Choi, Kim,
  et~al\mbox{.}}{Son et~al\mbox{.}}{2015}]%
        {son2015rocking}
\bibfield{author}{\bibinfo{person}{Yunmok Son}, \bibinfo{person}{Hocheol Shin},
  \bibinfo{person}{Dongkwan Kim}, \bibinfo{person}{Young-Seok Park},
  \bibinfo{person}{Juhwan Noh}, \bibinfo{person}{Kibum Choi},
  \bibinfo{person}{Jungwoo Choi}, \bibinfo{person}{Yongdae Kim},
  {et~al\mbox{.}}} \bibinfo{year}{2015}\natexlab{}.
\newblock \showarticletitle{Rocking drones with intentional sound noise on
  gyroscopic sensors}. In \bibinfo{booktitle}{{\em Proceedings of the USENIX
  Security Symposium}}. \bibinfo{pages}{881--896}.
\newblock


\bibitem[\protect\citeauthoryear{Sound}{Sound}{2017}]%
        {crysound}
\bibfield{author}{\bibinfo{person}{Cry Sound}.}
  \bibinfo{year}{2017}\natexlab{}.
\newblock \bibinfo{title}{{CRY343 free field measurment microphone}}.
\newblock
  \bibinfo{howpublished}{\url{http://www.crysound.com/product_info.php?4/35/63}}.
    (\bibinfo{year}{2017}).
\newblock


\bibitem[\protect\citeauthoryear{Speech}{Speech}{2017}]%
        {Selvy}
\bibfield{author}{\bibinfo{person}{Selvy Speech}.}
  \bibinfo{year}{2017}\natexlab{}.
\newblock \bibinfo{title}{{Demo-Selvy TTS}}.
\newblock
  \bibinfo{howpublished}{\url{http://speech.selvasai.com/en/text-to-speech-demonstration.php}}.
    (\bibinfo{year}{2017}).
\newblock


\bibitem[\protect\citeauthoryear{STMicroelectronics}{STMicroelectronics}{2014}]%
        {STMicroelectronics1}
\bibfield{author}{\bibinfo{person}{STMicroelectronics}.}
  \bibinfo{year}{2014}\natexlab{}.
\newblock \bibinfo{title}{{MP23AB02BTR MEMS audio sensor, high-performance
  analog bottom-port microphone}}.
\newblock
  \bibinfo{howpublished}{\url{http://www.mouser.com/ds/2/389/mp23ab02b-955093.pdf}}.
    (\bibinfo{year}{2014}).
\newblock


\bibitem[\protect\citeauthoryear{STMicroelectronics}{STMicroelectronics}{2016}]%
        {STMicroelectronics2}
\bibfield{author}{\bibinfo{person}{STMicroelectronics}.}
  \bibinfo{year}{2016}\natexlab{}.
\newblock \bibinfo{title}{{MP34DB02 MEMS audio sensor omnidirectional digital
  microphone}}.
\newblock
  \bibinfo{howpublished}{\url{http://www.mouser.com/ds/2/389/mp34db02-955149.pdf}}.
    (\bibinfo{year}{2016}).
\newblock


\bibitem[\protect\citeauthoryear{STMicroelectronics}{STMicroelectronics}{2017}]%
        {MEMSmicrophone}
\bibfield{author}{\bibinfo{person}{STMicroelectronics}.}
  \bibinfo{year}{2017}\natexlab{}.
\newblock \bibinfo{title}{{Tutorial for MEMS microphones}}.
\newblock
  \bibinfo{howpublished}{\url{http://www.st.com/content/ccc/resource/technical/document/application_note/46/0b/3e/74/cf/fb/4b/13/DM00103199.pdf/files/DM00103199.pdf/jcr:content/translations/en.DM00103199.pdf}}.
    (\bibinfo{year}{2017}).
\newblock


\bibitem[\protect\citeauthoryear{Sun, Jin, Chen, Zhang, Zhang, and Zhang}{Sun
  et~al\mbox{.}}{2016}]%
        {sun2016visible}
\bibfield{author}{\bibinfo{person}{Jingchao Sun}, \bibinfo{person}{Xiaocong
  Jin}, \bibinfo{person}{Yimin Chen}, \bibinfo{person}{Jinxue Zhang},
  \bibinfo{person}{Yanchao Zhang}, {and} \bibinfo{person}{Rui Zhang}.}
  \bibinfo{year}{2016}\natexlab{}.
\newblock \showarticletitle{VISIBLE: Video-Assisted keystroke inference from
  tablet backside motion}. In \bibinfo{booktitle}{{\em Proceedings of the
  Network and Distributed System Security Symposium (NDSS)}}.
\newblock


\bibitem[\protect\citeauthoryear{Technologies}{Technologies}{2017a}]%
        {jinci}
\bibfield{author}{\bibinfo{person}{Jinci Technologies}.}
  \bibinfo{year}{2017}\natexlab{a}.
\newblock \bibinfo{title}{Open structure product review}.
\newblock \bibinfo{howpublished}{\url{http://www.jinci.cn/en/goods/112.html}}.
   (\bibinfo{year}{2017}).
\newblock


\bibitem[\protect\citeauthoryear{Technologies}{Technologies}{2017b}]%
        {keysight}
\bibfield{author}{\bibinfo{person}{Keysight Technologies}.}
  \bibinfo{year}{2017}\natexlab{b}.
\newblock \bibinfo{title}{{N5172B EXG X-Series RF Vector Signal Generator, 9
  kHz to 6 GHz}}.
\newblock
  \bibinfo{howpublished}{\url{http://www.keysight.com/en/pdx-x201910-pn-N5172B}}.
    (\bibinfo{year}{2017}).
\newblock


\bibitem[\protect\citeauthoryear{to~Speech}{to~Speech}{2017}]%
        {fromtexttospeech}
\bibfield{author}{\bibinfo{person}{From~Text to Speech}.}
  \bibinfo{year}{2017}\natexlab{}.
\newblock \bibinfo{title}{{Free online TTS service}}.
\newblock \bibinfo{howpublished}{\url{http://www.fromtexttospeech.com/}}.
  (\bibinfo{year}{2017}).
\newblock


\bibitem[\protect\citeauthoryear{to~Speech~Technologies}{to~Speech~Technologies}{2017}]%
        {innoetics}
\bibfield{author}{\bibinfo{person}{Innoetics~Text to Speech~Technologies}.}
  \bibinfo{year}{2017}\natexlab{}.
\newblock \bibinfo{title}{{Innoetics Text-to-Speech}}.
\newblock \bibinfo{howpublished}{\url{https://www.innoetics.com/}}.
  (\bibinfo{year}{2017}).
\newblock


\bibitem[\protect\citeauthoryear{Trippel, Weisse, Xu, Honeyman, and Fu}{Trippel
  et~al\mbox{.}}{2017}]%
        {trippel2017walnut}
\bibfield{author}{\bibinfo{person}{Timothy Trippel}, \bibinfo{person}{Ofir
  Weisse}, \bibinfo{person}{Wenyuan Xu}, \bibinfo{person}{Peter Honeyman},
  {and} \bibinfo{person}{Kevin Fu}.} \bibinfo{year}{2017}\natexlab{}.
\newblock \showarticletitle{WALNUT: Waging doubt on the integrity of mems
  accelerometers with acoustic injection attacks}. In \bibinfo{booktitle}{{\em
  Proceedings of the IEEE European Symposium on Security and Privacy
  (EuroS\&P)}}. IEEE, \bibinfo{pages}{3--18}.
\newblock


\bibitem[\protect\citeauthoryear{Vaidya, Zhang, Sherr, and Shields}{Vaidya
  et~al\mbox{.}}{2015}]%
        {vaidya2015cocaine}
\bibfield{author}{\bibinfo{person}{Tavish Vaidya}, \bibinfo{person}{Yuankai
  Zhang}, \bibinfo{person}{Micah Sherr}, {and} \bibinfo{person}{Clay Shields}.}
  \bibinfo{year}{2015}\natexlab{}.
\newblock \showarticletitle{Cocaine Noodles: Exploiting the gap between human
  and machine speech recognition}. In \bibinfo{booktitle}{{\em Proceedings of
  the USENIX Workshop on Offensive Technologies (WOOT)}}.
  \bibinfo{publisher}{USENIX Association}.
\newblock


\bibitem[\protect\citeauthoryear{Viikki and Laurila}{Viikki and
  Laurila}{1998}]%
        {viikki1998cepstral}
\bibfield{author}{\bibinfo{person}{Olli Viikki} {and} \bibinfo{person}{Kari
  Laurila}.} \bibinfo{year}{1998}\natexlab{}.
\newblock \showarticletitle{Cepstral domain segmental feature vector
  normalization for noise robust speech recognition}.
\newblock \bibinfo{journal}{{\em Speech Communication\/}} \bibinfo{volume}{25},
  \bibinfo{number}{1} (\bibinfo{year}{1998}), \bibinfo{pages}{133--147}.
\newblock


\bibitem[\protect\citeauthoryear{Vocalware}{Vocalware}{2017}]%
        {vocalware}
\bibfield{author}{\bibinfo{person}{Vocalware}.}
  \bibinfo{year}{2017}\natexlab{}.
\newblock \bibinfo{title}{{Vocalware TTS}}.
\newblock \bibinfo{howpublished}{\url{https://www.vocalware.com/}}.
  (\bibinfo{year}{2017}).
\newblock


\bibitem[\protect\citeauthoryear{Wang, Wu, and Xu}{Wang et~al\mbox{.}}{2016}]%
        {wang2016windcompass}
\bibfield{author}{\bibinfo{person}{Xiaohui Wang}, \bibinfo{person}{Yanjing Wu},
  {and} \bibinfo{person}{Wenyuan Xu}.} \bibinfo{year}{2016}\natexlab{}.
\newblock \showarticletitle{WindCompass: Determine Wind Direction Using
  Smartphones}. In \bibinfo{booktitle}{{\em Proceedings of the 13th Annual IEEE
  International Conference on Sensing, Communication, and Networking (SECON)}}.
  IEEE, \bibinfo{pages}{1--9}.
\newblock


\bibitem[\protect\citeauthoryear{Xdadevelopers}{Xdadevelopers}{2017}]%
        {HiVoice}
\bibfield{author}{\bibinfo{person}{Xdadevelopers}.}
  \bibinfo{year}{2017}\natexlab{}.
\newblock \bibinfo{title}{{HiVoice app, what is it for?}}
\newblock
  \bibinfo{howpublished}{\url{https://forum.xda-developers.com/honor-7/general/hivoice-app-t3322763}}.
    (\bibinfo{year}{2017}).
\newblock


\bibitem[\protect\citeauthoryear{Yan, Xu, and Liu}{Yan et~al\mbox{.}}{2016}]%
        {yan2016can}
\bibfield{author}{\bibinfo{person}{Chen Yan}, \bibinfo{person}{Wenyuan Xu},
  {and} \bibinfo{person}{Jianhao Liu}.} \bibinfo{year}{2016}\natexlab{}.
\newblock \showarticletitle{Can you trust autonomous vehicles: Contactless
  attacks against sensors of self-driving vehicle}.
\newblock \bibinfo{journal}{{\em DEF CON\/}} (\bibinfo{year}{2016}).
\newblock


\end{thebibliography}

\end{document}